\documentclass[11pt,A4,twoside]{article}
\usepackage[hmargin=1.4in,vmargin=1.2in]{geometry} 
\usepackage{afterpage}
\usepackage{empheq}

\usepackage[deletedmarkup=xout]{changes}

\usepackage{natbib}\def\citeN{\citet*}\def\citeN{\citet*}

\usepackage{dsfont}

\usepackage[english]{babel}
\usepackage{enumerate,hyperref}
\usepackage{enumitem}
\usepackage{amssymb}
\usepackage{amsfonts}
\usepackage{booktabs} %
\usepackage{graphicx}
\usepackage{amsmath}
\usepackage{latexsym}
\usepackage{amsthm}
\usepackage{appendix}
\usepackage[all]{xy}
\usepackage{todonotes} %
\usepackage{fancyhdr}
\usepackage{comment}

\usepackage[linesnumbered,vlined,boxed,commentsnumbered]{algorithm2e}
\usepackage{caption}
\usepackage{subcaption}

\usepackage{breqn}
\usepackage{mathtools} %
\mathtoolsset{showonlyrefs=true} %
\usepackage{dsfont}
\usepackage{xspace}

\usepackage{placeins}
\let\oldsection\section
\renewcommand{\section}{\FloatBarrier\oldsection}
\let\oldsubsection\subsection
\renewcommand{\subsection}{\FloatBarrier\oldsubsection}
\let\oldsubsubsection\subsubsection
\renewcommand{\subsubsection}{\FloatBarrier\oldsubsubsection}

\usepackage{xr}%

\makeatletter
\def\timenow{\@tempcnta\time
\@tempcntb\@tempcnta
\divide\@tempcntb60
\ifnum10>\@tempcntb0\fi\number\@tempcntb
:\multiply\@tempcntb60
\advance\@tempcnta-\@tempcntb
\ifnum10>\@tempcnta0\fi\number\@tempcnta}
\makeatother

\numberwithin{equation}{section}

\newcommand \replytoreferee[2]{
\noindent\colorbox{red!5}{\begin{minipage}{\textwidth}
\color{red!100} 
\begin{description}
\item 
{\it Referee remark.}\\ #1
\item {\it Authors reply.}\\ #2
\end{description}
\end{minipage}}
}
\renewcommand \replytoreferee[2]{}

\newcounter{todocounter}
\newcommand{\emm}[1]{\stepcounter{todocounter}\todo[backgroundcolor=red!20!white,size=\tiny]{EG-\thetodocounter: #1}}\renewcommand{\emm}[1]{}
\newcommand{\emmi}[1]{\stepcounter{todocounter}\todo[inline,backgroundcolor=red!20!white,size=\small]{EG-\thetodocounter: #1}}
\renewcommand{\emmi}[1]{}
\definecolor{darkblue}{rgb}{0,0,.4}

\newcommand{\dav}[1]{\todo[inline,color=blue!10, author=David]{#1}}\renewcommand{\dav}[1]{}

\newcommand{\cA}{\mathcal{A}}
\newcommand{\cB}{\mathcal{B}}
\newcommand{\cF}{\mathcal{F}}
\newcommand{\cZ}{\mathcal{Z}}
\newcommand{\cFab}{\mathcal{F}^{a,b}}\renewcommand{\cFab}{\widetilde{\mathcal{F}}}\renewcommand{\cFab}{\mathcal{F}}

\newcommand{\cGB}{\mathcal{G}^B}\renewcommand{\cGB}{\widetilde{\mathcal{G}}}\renewcommand{\cGB}{\mathcal{G}}
\newcommand{\cH}{\mathcal{H}}
\newcommand{\cI}{\mathcal{I}}
\newcommand{\cL}{\mathcal{L}}

\newcommand{\cN}{\mathcal{N}}
\newcommand{\cR}{\mathcal{R}}

\newcommand{\cU}{\mathcal{U}}
\newcommand{\cX}{\mathcal{X}}

\newcommand{\N}{\mathbb{N}}
\newcommand{\R}{\mathbb{R}} 
\newcommand{\un}{\ind}
\newcommand\ind{\mathds{1}}

\newcommand \PE{\mathbb{E}}
\newcommand \ave[1]{\frac{1}{#1}\sum_{i=1}^{#1}}
\newcommand \Esp[1]{\mathbb{E}\left[#1\right]}

\newcommand \PP{\mathbb{P}}
\newcommand \PPro[1]{\PP\left[#1\right]}

\newcommand \ES{ s %
}

\newcommand \Radcemp{\cR_{emp}}
\newcommand \Radcave{\cR_{ave}}

\newcommand \objfunvar{\mathbf{ \VaR} %
}\renewcommand \objfunvar{q}
\newcommand \objfunexpsho{ \Delta_{{\mathbf \ES},\objfunvar} %
}
\renewcommand \objfunexpsho{r}
\newcommand\appvar{\widetilde{\objfunvar}^{{a,b}}}\renewcommand\appvar{\widetilde{\objfunvar}}

\newcommand \empappvar{\eg{\widehat{\objfunvar}^{{a,b}}}}\renewcommand \empappvar{\eg{\widehat{\objfunvar}} }

\newcommand \empappexpsho[1]{\eg{\widehat{\objfunexpsho}_{#1}}}%

\newcommand \empappexpshoB[1]{\eg{\widehat{\objfunexpsho}_{#1}^{{\;B}}}}\renewcommand \empappexpshoB[1]{\eg{\widehat{\objfunexpsho}_{#1}}}

\newcommand \empappes[1]{\eg{\widehat{s}_{#1}}}%

\newcommand\VaR{q%
}

\newcommand\VAR{\mathrm{VaR}}

\newcommand\lpvar{\Lambda}\renewcommand\lpvar{\Phi}\renewcommand\lpvar{p}\renewcommand\lpvar{\ell}\renewcommand\lpvar{\phi}
\newcommand{\lmeavar}{\widetilde{\lpvar}}\renewcommand{\lmeavar}{\Phi^{a,b}}\renewcommand{\lmeavar}{\Phi}

\newcommand \lempestrunc{\eg{\widehat{\lpes}^{{\;B}}}}\renewcommand \lempestrunc{\eg{\widehat{\Psi}^{{\;B}}}}\renewcommand \lempestrunc{\eg{\widehat{\Psi}}}

\def\gqh{\gamma_{\empappvar}}
\def\gqf{\gamma_{f}}

\def\proof{\noindent {\it Proof. $\, $}}\def\proof{\noindent{\it\textbf{Proof}. $\, $}}

\usepackage{amsthm}
\newtheorem{theorem}{Theorem}[section]%
\newtheorem{pro}[theorem]{Proposition} 
\newtheorem{cor}[theorem]{Corollary}
\newtheorem{lem}[theorem]{Lemma}

\newcommand{\bt}{\begin{theorem}}\newcommand{\et}{\end{theorem}}
\newcommand{\bl}{\begin{lem}}\newcommand{\el}{\end{lem}}
\newcommand{\bp}{\begin{pro}}\newcommand{\ep}{\end{pro}}
\newcommand{\bcor}{\begin{cor}}\newcommand{\ecor}{\end{cor}}
\newcommand{\bconj}{\begin{conj}}\newcommand{\econj}{\end{conj}}

\theoremstyle{definition}
\newtheorem{defi}{Definition}[section] 
\newtheorem{hyp}[defi]{Assumption}

\theoremstyle{remark}
\newtheorem{rem}{Remark}[section] 
\newtheorem{ex}[rem]{Example}
 \newtheorem{nota}[rem]{Notation}

\newcommand{\bd}{\begin{defi} %
}\newcommand{\ed}{\end{defi} }
\newcommand{\bn}{\begin{nota} %
}\newcommand{\en}{\end{nota} }
\newcommand{\brem }{\begin{rem} %
}\newcommand{\erem }{\end{rem}}
\newcommand{\bcom }{\begin{com} %
}\newcommand{\ecom }{\end{com}}
\newcommand{\brems }{\begin{rem} %
}\newcommand{\erems }{\end{rem}}
\newcommand{\bex}{\begin{ex} %
}\newcommand{\eex}{\end{ex}}
\newcommand{\bexo}{\begin{exo} %
}\newcommand{\eexo}{\end{exo}}

\def\bhyp{\begin{hyp} %
}\def\ehyp{\end{hyp}}

\def\eds{\ed}
\def\eexs{\eex}

\def\sp{\,,\;}

\renewcommand \PE{{\rm E}}
\renewcommand\ES{{\rm ES}}
\renewcommand\VaR{{\rm VaR}}

\def\qq{\quad}
\def\qqq{\quad\quad\quad}
\def\hun{h_1}\def\hun{\iota}\def\hun{}

\def\hde{h_2}\def\hde{\varsigma}\def\hde{}

\def\bb{\begin{block}}
\def\eb{\end{block}} 
\def\cNN{\mathcal{NN}}

\renewcommand \PE{{\rm E}}
\renewcommand \Esp[1]{\PE\left[#1\right]}

\renewcommand \PP{{\rm P}}
\renewcommand \PPro[1]{\PP\left[#1\right]}

\def\purp{}
\DeclareMathOperator*{\argmin}{\arg\!\min}

\def\theq{z_1}\def\theq{q}
\def\thes{z_2}\def\thes{s}
\def\ther{z}\def\ther{r}

\newcommand\eg[1]{{\color{black}#1}}

\newcommand\db[1]{{\color{green}#1}}
\renewcommand{\db}[1]{}

\def\finproof{\rule{4pt}{6pt}}\def\finproof{\ensuremath{\square}}
\def\fe{~\finproof}\def\fe{}

\def\VaRa{\VaR_{\alpha}}\def\VaRa{\VaR}
\def\ESa{\ES_{\alpha}}\def\ESa{\ES}
\def\qalpha{q_{\alpha}}\def\qalpha{q}
\def\salpha{s_{\alpha}}\def\salpha{s}

 \newcommand{\beqa}{\begin{eqnarray}}
\newcommand{\eeqa}{\end{eqnarray}}
\newcommand\bal{\begin{aligned}}
\newcommand\eal{\end{aligned}}
\newcommand{\beql}[1]{\begin{equation}\label{#1}\bal}
\newcommand{\eeql}{\eal\end{equation}}
\newcommand{\bel}{\begin{equation*}\bal}
\newcommand{\eel}{\eal\end{equation*}}

\def\lay{n}\def\lay{l}
\def\then{n}
\def\thenu{n}

\hypersetup{
 colorlinks  = true, %
 urlcolor   = black, %
 linkcolor  = black, %
 citecolor  = black %
}
\makeatletter
\renewcommand\@makefnmark{\hbox{\@textsuperscript{\normalfont\color{blue}\@thefnmark}}}
\renewcommand\@makefntext[1]{%
 \parindent 1em\noindent
      \hb@xt@1.8em{%
        \hss\@textsuperscript{\normalfont\@thefnmark}}#1}
\makeatother 

\def\Batches{B}\def\Batches{\Pi}

\def\otimes{\times}

\def\theu{v}\def\theu{u}
\def\thev{z}\def\thev{v}
\newcommand \theP{\eg{\cal Z}}

\def\pin{p}\def\pin{l}\def\pin{\ell}
\newcommand{\eqdef}{\mathrel{\mathop:}=}
\def\eqd{=}
\newcommand{\defeq}{\mathrel{{=}{\mathop:}}}
\def\nnx{({\cX})}\def\nnx{}
  \def\nn{1\,..\,n}\def\nn{1 .. n}\def\nn{1:n}
\def\Tbb{T^{B}} \def\Tbb{B\wedge}
\def\Tb#1{\Tbb\left({#1}\right)}

\def\lpesB{l^{B}}\def\lpesB{\psi^{B}}
\def\therB{\widetilde{\ther}}\def\therB{\ther^{B}}

\newcommand \den{\pi}\renewcommand \den{\dot{F}}

\newcommand{\F}{\tilde{\cF}}\renewcommand{\F}{\cF}

\def\Upsilon{F}

\DeclareSymbolFont{yhlargesymbols}{OMX}{yhex}{m}{n}
\DeclareMathAccent{\wideparen}{\mathord}{yhlargesymbols}{"F3}

\usepackage{threeparttable}

\def\cLp {\mathcal{G}}\def\cLp {\cL_+^2}
\def\cLab{\mathcal{L}_{a,b}}
\def\cLabt{\mathcal{L}_{a,b}}
\begin{document}

\title{Statistical Learning of\\ Value-at-Risk and Expected Shortfall\footnote{{
Python notebooks reproducing the results of this paper are available  
on \url{https://github.com/BouazzaSE/Learning-VaR-and-ES}. 
}
}
 }
\author{
D. Barrera\thanks{Email: {\tt j.barrerac@uniandes.edu.co}. Departamento de Matem\'{a}ticas, Universidad de los Andes, Cra 1 \# 18a-12, Edificio H. Bogot\'{a}, Colombia. Postal code: 111711. The research of D. Barrera
benefited from the support of the \textit{Chair Capital Markets Tomorrow: Modeling and Computational Issues} under the aegis of the Institut Europlace de Finance,  a joint initiative of Laboratoire de Probabilit\'es, Statistique et Mod\'elisation (LPSM) / Universit\'e Paris Cit\'e and  Cr\'edit Agricole CIB, and from the support of the  \textit{Chair Stress Test, RISK Management and Financial Steering of the Foundation Ecole Polytechnique.}}, 
S. Cr\'epey\thanks{Email: {\tt stephane.crepey@lpsm.paris}. LPSM, Universit\'e Paris Cit\'e,
France. The research of S. Cr\'epey benefited from the support of the \textit{Chair Stress Test, RISK Management and Financial Steering}, led by the French Ecole polytechnique and its Foundation and sponsored by BNP Paribas.} , 
E. Gobet\thanks{Email: {\tt emmanuel.gobet@polytechnique.edu}. 
Centre de Math\'ematiques Appliqu\'ees (CMAP), CNRS, Ecole Polytechnique, Institut Polytechnique de Paris. Route de Saclay, 91128 Palaiseau Cedex, 
France. The research of E. Gobet
is supported by the  \textit{Chair Stress Test, RISK Management and Financial Steering of the Foundation Ecole Polytechnique.}} , 
Hoang-Dung Nguyen\thanks{Email: {\tt hdnguyen@lpsm.paris}. PhD student, Universit\'e Paris Cit\'e, 
France. The research of H.-D. Nguyen is funded by a CIFRE grant from Natixis.} , 
B. Saadeddine\thanks{Email: {\tt bouazza.saadeddine2@ca-cib.com}. Quantitative research GMD, Credit Agricole CIB, Paris.}  
}

\date{\today}

\maketitle
 \begin{abstract} 
We propose  
a non-asymptotic convergence analysis of a two-step approach to learn a conditional value-at-risk (VaR) and a conditional expected shortfall (ES) using Rademacher bounds, in a non-parametric setup allowing for heavy-tails on the financial loss.
Our approach for the VaR is extended to the problem of learning at once multiple VaRs corresponding to different quantile levels. This results in efficient learning schemes based on neural network quantile and least-squares regressions. An a posteriori Monte Carlo procedure is introduced to estimate distances to the ground-truth VaR and ES. This is illustrated by numerical experiments in a Student-$t$ toy model and a financial case study where the objective is to learn a dynamic initial margin.

\medskip



\end{abstract}
 
\emmi{I put some replies to referee questions directly in the latex, to facilitate the revision. Of course, at the end, we will produce a separate reply letter collecting all these replies. The latex command is \tt $\backslash$replytoreferee$\{question...\}\{reply...\}$}

\section{Introduction\label{s:intro}}

\paragraph{Motivation.}
\purp{Quantile regression} is a classical statistical problem that has received attention since the 1750s. 
According to \citeN{koenker2017quantile}, the least absolute criterion (or pinball loss function) for the median even preceded the least squares for the mean (introduced by Legendre in 1805).  Quantile regression is commonly performed in the context of linear models, where the ensuing minimization problem can be cast as a linear program solved by the simplex method.  Alternative approaches include
nonlinear quantile regression based on interior point methods
\citep*{koenker1996interior} or
nonparametric quantile regression implemented by stochastic gradient descent \citep*{rodrigues2020beyond}.

 In harmony with the numerous financial applications, we refer to quantile as value-at-risk (VaR) and to superquantile \citep*{rockafellar2013superquantiles},  i.e.~the expected loss 
 given the loss exceeds the VaR, as expected shortfall (ES, initially named as conditional value-at-risk in the literature, however this terminology makes it confusing when considering additional conditioning as we do).
 In this paper we learn conditional versions of VaR and ES, accounting for some available information in the conditioning, represented by a random variable $X$. Conditional VaR and ES appear naturally in various financial applications. Learning the VaR and ES provides a way to 
shortcut nested Monte Carlo simulations by regression (cf.\ \citet*{BroadieDuMoallemi15} for learning or regressing a conditional expectation).  Our initial motivation was the simulation of dynamic initial margin 
or economic capital processes 
in the context of refined FVA or KVA computations as per
\citet*[Section A.4]{albanese2021xva} or \citet*[Section 5]{AbbasturkiCrepeyLiSaadeddine23}.
As a second application, one may consider a forward-looking risk management exercise,  whereby a bank samples several possible scenarios along which it must compute all its risk metrics in order to assess the amount of regulatory capital required to secure its activities.
A third application could be related to stress testing exercises required by regulators,
whereby the risk metrics need be evaluated along various stressed scenarios.
Importantly, our approach is readily extendible, both in theory and practice, to the problem of learning multiple quantiles at the same time, furnishing in particular efficient approaches to the so called crossing quantiles problem (see Section \ref{ss:qci} for a discussion of the related literature).
A fourth application of this work thus consists in parameterizing the VaR and ES as a function of their risk level $\alpha\in (0,1)$, so as to have an evaluation of risk metric functions conditionally on $\alpha$ (and other informative variables).
\replytoreferee{{\bf Motivation for considering the conditional version of VaR and ES.} The authors argued the contributions of learning conditional VaR and ES. It is not clear why this is important for financial applications. The numerical experiments are not enough to support a general scope of the framework. I understand that in terms of back-testing, sometimes it is important to calculate the losses of trading strategies conditioned on some rare events. The authors are encouraged to make a better case for the learning framework of conditional VaR and ES.}{We have added a paragraph in the introduction to expose the reasons why studying conditional risk measures is important.} 

\paragraph{Our contributions in relation to the literature.}
\citeN{dimitriadis2017} 
introduced a
 joint linear regression estimator for VaR and ES based on their joint elicitability properties \citep*{fiss:zieg:16,fiss:zieg:gnei:15}, 
implemented numerically using the 
nonlinear simplex 
optimization algorithm. 
They developed
an asymptotic convergence analysis for their estimators, establishing their consistency and asymptotic normality under somewhat strong semiparametric assumptions and  regularity conditions. Instead,
we propose a non-asymptotic convergence analysis of a learning algorithm for VaR
 and ES using a two-step approach in a nonparametric setup. 
 We then specify our results to 
 practical schemes for learning the conditional VaR and ES using neural networks as the function approximators.

\citeN{padilla2020quantile} also consider quantile regression with ReLU networks, but only in the case of deterministic covariates $X_i$ (in our notation). In a setup similar to our
Assumption \ref{assconfarvarcon} (see Remark \ref{r:padi}),
they provide qualitative non-asymptotic estimates for such networks, 
of which our corresponding results can be considered quantitative versions (i.e.\ with explicit constants); 
they also establish minimax rates for quantile functions with H\"{o}lder-related regularity or improved rates under Besov regularity on the target function.

Assuming that the quantile function has a compositional structure in terms of H\"older-continuous functions, \citeN{shen2021deep} derive non-asymptotic
error bounds that depend only  on the dimension of the composed functions (as opposed to the dimension $d$ of the inputs usually in the literature), but
a statistical error 
term requiring an order of integrability $p>1$ of the response variable $Y$. 

As opposed to the previous references who mainly follow Vapnik-Chervonenkis (VC) based approaches,
we mainly follow a Rademacher based approach.
Our value-at-risk error bounds do not require any finite moments beyond integrability of $Y$. In the neural network case, depending on the nature of the regularization that is used, our results may not depend on the dimension $d$ of the inputs either.
Also accounting for the values of the constants, fully explicit in our case, which appear in our error bounds,
we show in the  discussion following Theorem \ref{thebouerrrrad} how Rademacher-based bounds are better than VC bounds for data in the small to medium size regime that matters in finance, while VC only dominates Rademacher for really big data that may be relevant in other application fields.
Notice that the two papers discussed above only consider value-at-risk (quantile), as opposed to expected shortfall also in our case. Since the first arXiv version of this work was published, other works appeared on the two-step approach for learning the ES, notably \citet*{he2023robust}, sometimes with a non-asymptotic error analysis, but always restricted to a parametric or semiparametric linear setup.
\emmi{not yet done on my side, but this paragraph could be improved: We repeat several times that we evaluate multiple quantiles. In addition it is not clear enough how we should present the two cases, i.e.\ the usual conditional VaR/ES and the multiple quantiles. For me, multiple quantiles is a particular case of the global use-case
conditional VaR/ES, 
}

Beyond theory, we contribute by several algorithmic tricks.
  In the context of machine learning on simulated data, which is very relevant for quantitative finance modulo validability,
Proposition \ref{p:twin}  underpins an a posteriori error estimation method in order to compute errors against ground-truth values of the conditional VaR and ES (even without access to the latter).
 In the neural net setup, our two-step methodology enables the reuse of the VaR neural network's hidden layers in the training of the neural network approximating the ES, reducing the additional learning of the ES to a linear regression against the already learned regression basis for the VaR. 
 The multi-quantile learning approaches of Section \ref{sec:multi_alpha} (with the multi-$\alpha$(III) approach of Section \ref{par:var_interp_net} often found the best alternative in our numerics) not only diminish the computational burden with respect to several single-quantile learnings that would be run separately, but are also found to better learn the value-at-risks of high confidence levels $\alpha$.
\medskip

\paragraph{Outline of the paper.}
Section \ref{secprealg} introduces the setup and our learning algorithm.
Section \ref{algo:ca} delivers the corresponding convergence analysis.  
Section \ref{sec:learningwithneuralnets} discusses specializations of this scheme and its errors to the case of inference via neural networks. 
We introduce multi-quantile extensions of the above in Section~\ref{sec:multi_alpha}. Sections \ref{sec:toymodel} and \ref{sec:imstoprocess} discuss numerical experiments. 
Section \ref{s:qsrepr}  reviews classical 
properties of the unconditional VaR and ES. Technical proofs
are deferred to Section 
\ref{sec:tecpro}.

 \section{A Learning Algorithm for VaR and ES \label{secprealg}}

We denote by $(\Omega, \cA, \PP)$ a fixed probability space, 
 which admits all the random variables appearing below, with corresponding expectation operator denoted by $ \Esp{\,\cdot\,}$. 
      The notation $L^p_{\PP}$ (with $p\in [1,+\infty]$) stands for the usual \PP-integrability spaces and we denote by $\|.\|_{\PP,p}$ the related norms.
      For a general Polish space $\cZ$ with Borel sigma algebra ${\cB_{\cZ}}$, 
     by a random element $Z$ of ${\cZ}$ we mean an $\cA/{\cB_{\cZ}}$ measurable function $Z: \Omega\to {\cZ}$; 
      $\PP_{Z}$ denotes the law induced by $Z$ on $\cZ$, i.e.,
 for every $A\in \cB_{\cZ}$,  
$ 
\PP_{Z}(A)\eqd \PPro{Z\in A}.$ 

From now on, $\cX$  denotes a fixed Polish space
and
\begin{align}
\label{equhypxy}
(X,Y):\Omega\to {{\cX}}\times \R
 \mbox{\,\,\,is a fixed random element of ${\cX}\times \R$ with $Y\in L^{1}_{\PP}$.} 
 \end{align}  
We fix a conditional distribution function $
 \mu: {\cX}\times {\cB_\R}\to [0,1]$
of $Y$ given $X$ 
\citep*[Theorem 5.3 p.84]{kallenberg2006foundations} 
and we assume that the function $ {\cX}\times \R\to \R$ defined by $
(x,y)\mapsto \mu(x,(-\infty,y])$
is $({\cB_{\cX}}\otimes {\cB_\R})/{\cB_\R}$ (i.e.\ Borel) measurable (as for instance if ${\cX}=\R^{d}$ 
and $(X,Y)$ admits a density with respect to Lebesgue measure).
We  use  the corresponding version
$
\PPro{Y\in \cdot |X}\eqd \mu(X,\cdot)
$
of the conditional probability of $Y$ given $X$ and the conditional cdf of $Y$ given $X$,
$
 F_{Y|X}(y)\eqd \PPro{Y\le y \,|\, X}
 \eqd \mu(X,(-\infty,y]). 
 $
 We assume, without loss of generality, that $F_{Y|X(\omega)}(\cdot)$  
is integrable for every $\omega\in \Omega$ (since $Y\in L^{1}_{\PP}$, we have that
$
\infty>\Esp{|Y|}=\Esp{\Esp{|Y| | X}}=\Esp{\int_{\R}|y| F_{Y|X}(dy)},
$
thus $%
F_{Y|X(\omega)}
$ is integrable for $\PP$  a.e.\ $\omega$: it suffices to change the version of $X$
 to guarantee integrability for every $\omega$).
In particular, if $g:\R\to \R$ is such that $g(Y)$ is $\PP$  integrable, then
 \begin{align}
\label{equchacondisfunexp}
     \Esp{g(Y)|X}(\omega)= \int_{\R}g(y) F_{Y|X(\omega)}(dy),\qq \PP\mbox{ a.s.}. 
 \end{align}

For a function $F:\R\to \R$ and $q\in \R,$ 
$
F(q-)\eqdef \lim_{z \uparrow q } F(z).
$
\bd
\label{defqualevalp}
The conditional value-at-risk ($\VAR$) and expected shortfall (ES) of $Y$ given $X$ at the confidence level $\alpha\in(0,1)$ are
(cf.~\eqref{equdefvares})  
\begin{align}
&\VaRa (Y|X)\eqd \VaRa (F_{Y|X})={\min} F_{Y|X}^{-1}([\alpha,1])={\min}\{y\in\R:\, F_{Y|X}(y)\ge \alpha \},\\
& \ESa (Y|X)\eqd \frac{1}{1-F_{Y|X}(\VaRa (Y|X){-})}\int_{[\VaRa (Y|X),\infty)}y \,F_{Y|X}(dy).
\label{equdefvarxesxprel}
\end{align} 
\eds

\bl \label{corexifunvarexpsho} 
{There exist} Borel measurable functions $\qalpha: {\cX}\to \R$ and ${\salpha} : {\cX}\to \R$ such that
\begin{align}
\label{equdefvarxesx}
\qalpha(X)=\VaRa (Y|X)\sp  \salpha(X)=\ESa (Y|X),\qq \PP\mbox{ a.s.}.
\end{align}
\el

\proof  See Section \ref{secpromeavar}. \finproof

\bn \label{n:r} Besides \eqref{equdefvarxesx}, we also introduce $\ther=\thes-\theq,$ which is nonnegative, $\PP_X$ a.s..
\en

\bl \label{lemma:bounds:q}  {\rm\textbf{(i)}}  With probability one, we have 
\begin{align} 
    \label{bounds:on:VaR}
    -\alpha^{-1}\Esp{Y^-|X} &\leq q(X)
    \leq (1-\alpha)^{-1}\Esp{Y^+|X},\\
    \label{bounds:on:ES}
    |\thes(X)|&\leq (1-\alpha)^{-1}\Esp{|Y|\,|X}.
\end{align}
{\rm \textbf{(ii)}}  The (resp.\  square) integrability of $\theq(X)$ and $\thes(X)$ follows from that of $Y$.
\el
\proof See Section \ref{secprolembouq}. \finproof

\subsection{VaR and ES as Minimizers}

For our VaR error control, we assume the knowledge of a tube of bounded width 
containing the graph of the target function $q$. The more a priori knowledge one has on  $q$, i.e. the narrower this tube, the better will be the statistical error of our a priori estimates (see e.g.\ the second line in \eqref{equesterrvardel}).
\bhyp
\label{assconfarvarcon}
There exist 
functions $a,b:{\cX}\to \R$ such that \textbf{(i)}
the law $F_{Y|X}(\cdot)$ has a density $\den_{Y|X}(\cdot)$ on 
$[a(X),b(X)], \; \PP$  a.s.,
\textbf{(ii)} 
\begin{align}
\label{eququasan}
F_{Y|X}(a(X))< \alpha < F_{Y|X}(b(X)) \mbox{ holds $\PP$  a.s.,}
\end{align}
and 
\textbf{(iii)} $ 
0< \| b-a \|_{\PP_X, \infty}<\infty.$
\ehyp
\noindent

 
\brem \label{remassste}
The absolute continuity of $F_{Y|X}$ around the quantile of level $\alpha$ postulated in Assumption \ref{assconfarvarcon}(i) is a common assumption in the literature, see e.g.\ \citet[Chapter 3]{reis:12}. It can be ensured (with a quantifiable impact on the quantiles) by adding a small independent Gaussian noise to $Y$.
Assumptions \ref{assconfarvarcon}(i)-(ii) imply, in particular, that $F_{Y|X}(q(X))=\alpha$ and $a(X)<  q(X) < b(X)$ $\PP$ a.s.\ hold, but the converse is not true as there can be $\alpha$-quantile functions $b$ greater than the left-quantile (VaR) $q$. 
 Assumption \ref{assconfarvarcon}(iii) holds in particular if 
$Y$ is bounded or if $\cX$ is compact and $\theq$ can be chosen continuous, case in which $\theq$ is also bounded, but it may also hold when neither $X$ nor $Y$ (or $Y$ given $X$) nor $\theq$ are bounded.  
The $\cX$ compact and $q$ continuous case may seem restrictive   
but even this very special sub-case of Assumption \ref{assconfarvarcon} covers most use-cases of interest, as data  practitioners typically  restrict their learnings to compact sets of the state space, even if this means truncating it and doing several learnings on different subdomains.
Also note that one can always reduce the problem to the case of bounded $Y$ by applying a bijective increasing, hence VaR preserving (modulo an application of the inverse bijection to the estimated quantile function), bounded transformation to the data $Y$---with the caveat that applying such transformation and its inverse may not be innocuous numerically. In any case, the key point, which we see as a significant contribution of this work, is that Assumption \ref{assconfarvarcon} makes mainly (even though implicitly) assumptions on $X$, and no tail assumptions on $Y$ or $Y|X$ (beyond the integrability 
and square integrability
of $Y$ that are postulated in the VaR and ES respective parts of the paper).
This is important for applications in finance as \citet*{cont2022tail} demonstrate that heavy tail losses frequently arise from trading strategies, even in Gaussian models.
\erem

 \brem\label{r:padi} 
\citet[Assumption 2]{padilla2020quantile} can be seen as a slightly stronger version of Assumption   \ref{assconfarvarcon}, in the sense that the width of their guarantee tube on $q$ does not depend on $x$ and they assume positive lower and upper bounds on the density $\den_{Y|X}(\cdot)$ along their tube.
\citet{shen2021deep} even assumes $q$ ($f_0$ in their notation) bounded throughout the paper.
\erem



\bn\label{n:Datat}
Under Assumption \ref{assconfarvarcon}, $Y^{a,b}(\omega)\eqdef   a(X(\omega))\vee Y(\omega) \wedge  b(X(\omega)), \omega\in \Omega$, 
 and 
   $\cLab $ denotes the set of   functions $f$ on ${\cX}$
such that $f(x)\in [a(x),b(x)]$ (resp.\ $f(x)\geq 0$) holds for all $x$; 
$\cLp \nnx  $ denotes the set of $\PP_X$ square integrable functions $f$ on ${\cX}$
such that $f(x) \in [0,+\infty)$ holds for all $x$.
\en 

\bl\label{l:Linte} Under Assumption \ref{assconfarvarcon}, the (resp.\ square) integrability of $Y$ implies that any function $f\in \cLab $ is $\PP_X$ (resp.\ square) integrable.
\el
\proof Assumption \ref{assconfarvarcon}(i)-(ii) implies that
\beql{e:ab_inte}
&q(X)< b(X) = b(X) - a(X) + a(X) <
b(X) - a(X) + q(X)\\ 
&q(X) > a(X) =a(X)- b(X) + b(X) >a(X)- b(X) + q(X)
\eeql
hold $\PP$ a.s..\ By Lemma \ref{lemma:bounds:q}, the (resp.\ square) integrability of $Y$ implies the (resp.\ square) integrability of $q$. Combining this with the bounded $b-a$ in Assumption \ref{assconfarvarcon}(iii),  \eqref{e:ab_inte} yields the (resp.\ square) integrability of $a$ and $b$, which implies the one of any $f\in \cLab $. \finproof

\bd
\label{deflosfun}
Given $\alpha\in (0,1)$, the respective pointwise loss functions for  $\VaRa$ and ES$-$VaR given the hypothesis $u$ for $\VaRa$, both at the level $\alpha$, are given respectively by
\beql{equdefinslosvar}
&\mbox{the pinball loss, i.e. } \R^2\ni(y,\theu  )\mapsto \lpvar_{\hun}(y,\theu  ) = (1-\alpha)^{-1}(y-\theu  
)^{+}+ \theu \in\R  \sp \mbox{resp.}\\&
\R^2\times [0,+\infty) \ni (y,\theu,\thev  )\mapsto  ((1-\alpha)^{-1}(y-\theu  )^{+}-\thev  )^{2}=(\lpvar_{\hun}(y,\theu  ) -\theu  -\thev  )^{2}\in[0,+\infty).
\eeql
\ed
By application of Lemma \ref{lemma:bounds:q}(ii), the $\PP$  integrability of $Y$ also implies that of 
$\lpvar_{\hun}(Y,\theq(X))$; if $Y\in L_{\PP}^{2}$, then 
$(\lpvar_{\hun}(Y,\theq(X))-\theq(X)-r(X) )^{2}
$ is $\PP$  integrable. 
The loss functions $\lpvar(y,\theu  )$ and $(\lpvar_{\hun}(y,\theu  )-\theu-\thev  )^{2}$ underlie the following representations of
the functions $q$ and $r=s-q$ in Lemma \ref{corexifunvarexpsho} 
and Notation \ref{n:r} 
as solutions to optimization problems.
 
 \bl
 \label{thechareg}
 Under Assumption \ref{assconfarvarcon},  
 we have
\begin{align}
 \theq \in & \argmin  _{f\in\cLab\nnx  }\Esp{\lpvar_{\hun}(Y,f(X) )} =  \argmin  _{f\in\cLab\nnx  }\Esp{\lpvar_{\hun}(Y^{a,b},f(X) )}.
 \label{equcharminvarpoi}
 \end{align}
Moreover, any function belonging to either argmin in \eqref{equcharminvarpoi} is an $\alpha$-quantile function of $Y$ or, equivalently, of $Y^{a,b}$. In addition, for any function $\theq'$ in either argmin in \eqref{equcharminvarpoi}: 
\begin{align}
 \label{equexpshoconexp}
 \thes(X)=&\theq'(X)+(1-\alpha)^{-1}\Esp{(Y-q'(X))^{+}|X}\mbox{ holds } \PP\mbox{ a.s.};
 \end{align}
if $Y\in L_\PP^2$ and $q'\in\cLabt$, then 
 \begin{align}
 \thes -\theq'  \in & \argmin  _{g\in\cLp \nnx  } \Esp{(\lpvar_{\hun}(Y,\theq'(X)  )-\theq'(X)-g(X) )^{2} 
 } 
\label{equcharminexpshogivvarpoi} 
 \end{align} 
 and, for any function $\ther'$ belonging to the argmin in \eqref{equcharminexpshogivvarpoi}, 
 \begin{align}
     \thes(X)=\theq'(X)+\ther'(X)\mbox{ holds }\PP\mbox{ a.s.}.
 \end{align}
 \el
 \begin{proof} Using \eqref{equchavarmin} and \eqref{equchacondisfunexp},
 for every $f\in \cLab\nnx  \subset   L^{1}_{\PP_{X}} $ (by Lemma \ref{l:Linte}),
we obtain  
$
\Esp{\lpvar_{\hun}(Y,\theq(X) )|X}\leq \Esp{\lpvar_{\hun}(Y,f(X) )|X}   \PP\mbox{ a.s.},$
hence $\Esp{\lpvar_{\hun}(Y,\theq(X) ) }\leq \Esp{\lpvar_{\hun}(Y,f(X) ) }.$
This implies  that $q\in\argmin  _{f\in\cLab\nnx  }\Esp{\lpvar_{\hun}(Y,f(X) )}.$
Moreover, by \eqref{eququasan}, the conditional $\alpha$-quantiles of $Y^{a,b}$ and of $Y$ are the same; in particular
 \begin{align}
     \theq(X)=\VaR(Y^{a,b}|X),\qq \PP\mbox{ a.s.},
 \end{align}
and the second equality of \eqref{equchavarmin} 
 yields 
 \eqref{equcharminvarpoi}.

Conversely, if $g\in \cLab\nnx $
is  not an $\alpha$-quantile function of $Y$, i.e.\ if
$F_{Y|X}(g(X))$ differs from $\alpha$ on a set of positive $\PP$ probability,
then, according to \eqref{equchavarmin} and \eqref{equchacondisfunexp}, the random variable $Z=\Esp{\lpvar(Y,g(X))|X}-\Esp{\lpvar(Y,\theq(X))|X} $ satisfies
\begin{align}
&\PPro{Z\geq 0}=1 \mbox{ and }  \PPro{ Z>0}>0 ,
\end{align}
whence 
\begin{align}
   \Esp{ \lpvar(Y,g(X))}-\Esp{\lpvar(Y,\theq(X))}=  \Esp{Z}>0,
\end{align}
showing that $g$ does not belong to $\argmin  _{f\in\cLab \nnx  }\Esp{\lpvar_{\hun}(Y,f(X) )}$. Likewise,
  $$ \argmin _{f\in\cLab\nnx  }\Esp{\lpvar_{\hun}(Y^{a,b},f(X) )}$$  is included in the set
of the $\alpha$-quantile functions of $Y^{a,b}$ or, equivalently, of $Y$.

Finally,  \eqref{equchaexpshomin} and \eqref{equchacondisfunexp} yield \eqref{equexpshoconexp}, from which \eqref{equcharminexpshogivvarpoi} follows
by the representation of the conditional expectation as an $L^2$ projection in the square integrable  case. \finproof
 \end{proof}

 \brem
 \label{remnonuni}
 The minimizers in \eqref{equcharminvarpoi} do not need to be unique: any function $q'\in\cLab$ satisfying $F_{Y|X}(q'(X))=\alpha$ $\PP$  a.s.\ is a minimizer of $f\mapsto\Esp{\lpvar_{\hun}(Y,f(X) )}$ and there are infinitely many such functions $q'$ (and random variables $q'(X)$) in the ``degenerate case'' where $F_{Y|X}^{-1}(\alpha)$ is an interval of positive length on a set of positive  $\PP_{X}$  measure.
 \erems

  \brem
 \label{remtailgan}
There exist whole families, generalizing  \eqref{equcharminvarpoi} and \eqref{equcharminexpshogivvarpoi}, of representations of the functions $q$ and $r=s-q$ (or $q$ and $s$) as minimizers of suitable functionals, including joint 
(but non globally convex) representations of the pair $(q,s)$ based on the joint elicitability properties of value-at-risk and expected shortfall \citep{fiss:zieg:16,fiss:zieg:gnei:15}: see Theorem 2.3 in the arXiv v1 version of this work.
Using such a joint representation of $(q,s)$ was actually our original choice for practical computations in 
\cite{albanese2021xva}, in the footsteps of  \cite{dimitriadis2017}.
Deriving the VaR and ES at once indeed looks an attractive idea, but after more empirical investigation reported in the paper's GitHub  the best turned out to be the simplest, i.e.\ the two-step algorithm that first produces an approximation of the (conditional)  VaR and then learns the ES by least-squares using the VaR approximation. 

The joint approach in fact suffers from two different scales present in the same problem, to the effect that either the VaR or the ES is badly handled. Moreover, the mathematical analysis of the joint approach poses difficulties of its own, even under locally convex parametrizations such as the ones considered (for 
tail risk scenario generation) in \cite{cont2022tail}; also note that (local or global) convexity at the functional level does not  necessarily imply convexity with respect to the weights of a neural net approximator. 

In view of these practical and theoretical considerations,  we focus on the two-step algorithm hereafter. On this two-step approach (in a semiparametric linear setup), see also \citet*{he2023robust}.
 \erems

\noindent{\bf The two-step learning scheme for VaR and ES.}
The functional representations \eqref{equcharminvarpoi} and  \eqref{equcharminexpshogivvarpoi} give rise to associated approximation algorithms for $q$ and $s$.
The numerical recipe is just to replace the minimization problems in \eqref{equcharminvarpoi} or \eqref{equcharminexpshogivvarpoi}  by empirical versions: instead of $\cLab\nnx  $ and $\cLp \nnx  $, we use convenient hypotheses spaces (families of functions represented by neural nets in our numerics) inside the previous ones; instead of $\PP$  expectation, we use a Monte Carlo approximation based on 
 i.i.d.\ samples $(X,Y)_{\nn}  
=
((X_1,Y_1),\dots,(X_n,Y_n))$ of $(X,Y)$ in $\cX\times \R$, with $(X,Y)$ independent of the sample.

\section{%
 Convergence Analysis
\label{algo:ca}}
 
\def\Yab{Y^{a,b}}

The a priori error analysis of Sections \ref{ss:vars}--\ref{ss:esv}
corresponds essentially to the above scheme based on the data $(X,Y)_{\nn} $, ignoring the numerical optimization error (i.e.\ assuming one has access to global minimizers of the empirical risk functions). A more practical (and exhaustive, but only a posteriori) error control is then provided in Section \ref{ss:trick}.
 

\subsection{Estimation of VaR: General Setting}\label{ss:vars}

For a possibly data dependent  $f\in \cLab\nnx  \subset   L^{1}_{\PP_{X}} $ (by Lemma \ref{l:Linte}),
i.e.\ a function of $\cLab$ possibly parameterized by the data $(X,Y)_{\nn},$  we define
\bel&
\lmeavar_{\hun}
(f)\eqd \int_{\cX\times\R}\lpvar_{\hun}\Big(a(x)\vee y\wedge b(x) ,f(x )\Big)  \PP_{X,Y}(dx,dy)=
\Esp{\lpvar_{\hun}\left(\Yab ,f(X )\right)  \Big| (X,Y)_{\nn}   },
\eel
(cf.\ Notation \ref{n:Datat}).
We also fix a hypothesis space 
$
\cFab\subseteq \cLab\nnx       
$
ensuring the existence of
\begin{align}
  \appvar\in \argmin _{f\in \cFab}\lmeavar_{\hun}(f)\mbox{ and } \empappvar\in \argmin  _{f\in \cFab} \ave{n}\lpvar_{\hun}\left(\Yab_i,f(X_{i}) \right)    
  \label{equappvarprel}
  ,
  \end{align}
   dubbed best mean and best empirical hypothesis for VaR within $\cFab$.

\subsection{A Priori Error Bounds for the Estimation of VaR %
}
\label{secapperrvar}

Under Assumption \ref{assconfarvarcon}, $\theq$ as in \eqref{equdefvarxesx} defines an $\alpha$-quantile function (the smallest one for $\PP_{X}$  a.e.\ $x$)
. For such $\theq$, define 
\begin{align}
\label{equdefweifuncas}
\gamma_{f}(x)\eqd \Gamma_{{F_{Y|X=x}}}(f(x),\theq(x))
\end{align}
via \eqref{equdefgammf}, 
so that 
$\gamma_{f}(x)$ multiplied by $2(1-\alpha)$ coincides with $\den_{Y|X=x}$ evaluated at some middle point between $f(x)$ and $\theq(x)$ (see \eqref{equequlamtay}).
The following lemma interprets the approximation error  $\lmeavar(\appvar)-\lmeavar(\theq)$ as a weighted distance between $ \appvar $ and $\theq$ in the $L^{2}$ norm; by \eqref{equirrwei}, with $\PP_{X}$  probability one, the weights $\gamma_{f}(x)$ are equal to zero at some $x$ only if $f(x) $  is already an $\alpha$-quantile of $F_{Y|X=x}$.
Note that the inequality in \eqref{equineapperrvar} is similar to Lemma 3 in \cite{shen2021deep}.
\bl
 \label{theapprisexiden} {Under Assumption \ref{assconfarvarcon}, with 
 $\gqf$ as in \eqref{equdefweifuncas},  
 \begin{align}
 \label{equirrwei}
\PP\Big(\{\gqf(X)=0\}\setminus\{F_{Y|X}(f(X))=\alpha\}\Big)=0
 \end{align}
holds for any $f\in \cLab\nnx $. Moreover, for $\appvar$ defined  by \eqref{equappvarprel},
\beql{equineapperrvar}
\lmeavar(\appvar)-\lmeavar(\theq)=\|(\appvar-\objfunvar){\gamma_{\appvar}}\|_{\PP_{X},2}^{2}&= \inf_{f\in \cFab}\|(f-\objfunvar){\gamma_{f}}\|_{\PP_{X},2}^{2}   \leq %
\frac{2-\alpha}{1-\alpha}\inf_{f\in \cFab}\|f-\objfunvar\|_{\PP_{X},1}.
\eeql
}
 \el
 
\proof See Section \ref{secprolemuppboubia}. \finproof

\bex
\label{exauniappabs}
Assume that  
\beql{equremunibou}
0<c\leq \gamma_{\appvar}(X)
 \eeql
holds $\PP$ a.s.\ for some  constant $c$
(according to \eqref{equboudiflos} this is true for instance if 
\beql{e:cC} c\leq \frac{1}{2(1-\alpha)}\den _{Y|X}(t f(X)+(1-t)\theq(X))
\leq C \eeql
holds  $\PP$  a.s.\ for every $t\in (0,1)$ and $f\in\cF$, for some positive constants $c\le C$). Suppose additionally that
$
 \inf_{f\in  \cFab }  \|f -\objfunvar\|_{\PP_{X},1}<\delta 
$
holds for some positive constant $\delta$.
Then \eqref{equineapperrvar}
yields
 \beql{equesterruniapp}
{\sqrt{c}}\|\appvar-\objfunvar\|_{\PP_{X},2}\leq \sqrt{\frac{2-\alpha}{1-\alpha}\delta}.\fe
 \eeql
\eex

 \bex
 \label{exauniappcon}
Assume that  (i)   \eqref{equremunibou} holds, (ii) \eqref{eququasan} holds for some constants $a(X)\equiv A$ and $b(X)\equiv B$, i.e.\ $
 \objfunvar(X)\in (A,B), \PP\mbox{ a.s.},$
and (iii) there exists 
(e.g. because ${\cX}$ is compact and $\objfunvar$ H\"older continuous)
an  enumerable partition $\{{\cX_j}\}_{j}$ of measurable subsets of ${\cX}$ such that
$
\sup_{j} \sup_{(x,x')\in {\cX_j}\times {\cX_j}} |\objfunvar(x)-\objfunvar(x')|<\delta  .
$
Then \eqref{equesterruniapp} holds for the following hypothesis space $\cFab$ (over which $\appvar$ minimizes $\lmeavar$, cf.\ \eqref{equdefweifuncas})
 \begin{align*}
 \cFab  = \Big\{x\mapsto \sum_{j} c_{j}\textbf{1}_{{\cX_j}}(x); c_{j}\in [A,B], \forall j\Big\}.
 \end{align*} 
 \fe
 \eexs

We now give an upper bound for an error in probability associated with the empirical estimator $\empappvar$ of $\theq$. For this, we need to introduce the following measures of complexity applicable to our hypothesis spaces. 
\bd
\label{defradcom}
Let $\theP$ be a Polish space and
$\cH$ be a set of measurable real valued functions on $\theP$.
For any random sequence $Z_{\nn}  $  in $\theP$, the empirical Rademacher complexity $\Radcemp(\cH,Z _{\nn}  )$ and  the Rademacher complexity $\Radcave(\cH,Z _{\nn}  )$ of $\cH$ at $Z _{\nn}  $
 are defined as
\begin{align}
\Radcemp(\cH,Z _{\nn}  )=\Esp { \frac{1}{n} \sup_{h\in \cH}\sum_{k=1}^{n}U_{k}h(Z _{k}) \Big| Z _{\nn}  } 
\sp \Radcave(\cH,Z _{\nn}  )=\Esp{\Radcemp(\cH,Z _{\nn}  )} ,
\end{align}
where $U_{\nn}  $ is an i.i.d.\ Rademacher sequence $\PPro{U_{k}=1}=\PPro{U_{k}=-1}=1/2$ independent of $Z _{\nn}  $. 
\ed


\bt
\label{theeststaerrvar}
Under the assumptions of Lemma \ref{theapprisexiden}, for any $\delta\in (0,1)$, 
\beql{equesterrvardel}
&(1-\alpha)^{1/2}\left\|(\empappvar-\theq) \gqh\right\|_{_{\PP_{X},2}}
 \leq  
(1-\alpha)^{1/2}\inf_{f\in \cFab}\|(f-\objfunvar)\gqf\|_{\PP_{X},2}
 \\ 
 &\qqq\qqq+ \left(2(2-\alpha)\Radcave(\cFab,X_{\nn}  )+ 
\|b-a\|_{\PP_X,\infty} \sqrt{\frac{2\log (2/\delta)}{n}}\right)^{1/2}
\eeql
holds with probability at least $1-\delta$. 
\et
\proof See Section \ref{secprolemstaerrvar}. \finproof

\brems
From Theorem \ref{theeststaerrvar}, if there exist some constants $c,C$ such that
\beql{}
0<c\leq \gqf(X) \leq C <\infty\mbox{ holds }\PP\mbox{ a.s.},
\eeql
for every $f\in \cFab$ (e.g.\ if \eqref{e:cC} holds), then it can be deduced from \eqref{equesterrvardel} and \eqref{equineapperrvar} that  
\beql{equesterrvardel2}
&  c(1-\alpha)^{1/2} \left\|\empappvar-\theq \right\|_{_{\PP_{X},2}}
 \leq \left((2-\alpha)\inf_{f\in \cFab}\|f-\objfunvar\|_{\PP_{X},1}\right)^{1/2}\wedge 
 \left( C(1-\alpha)^{1/2}\inf_{f\in \cFab}\|f-\objfunvar\|_{\PP_{X},2}\right)
 \\ 
 &\qqq+ \left(2(2-\alpha)\Radcave(\cFab,X_{\nn}  )+ 
\|b-a\|_{\PP_X,\infty} \sqrt{\frac{2\log (2/\delta)}{n}}\right)^{1/2}
\eeql
holds with probability at least $1-\delta$. A better a priori knowledge on $q$, i.e.\ a smaller $\|b-a\|_{\PP_X,\infty}$ (see above Assumption \ref{assconfarvarcon}), also results into larger $c$ and smaller $C$, hence better constants in \eqref{equesterrvardel2}, but always with the same rate $n^{-\frac{1}{4}}$ for the statistical error assuming $\Radcave(\cFab,X_{\nn}  )={\rm O}( n^{-\frac{1}{2}} )$ as satisfied in our neural network application below.
\erems

\brem 
In the proof of Lemma \ref{l:Linte}, $\cLab\subset L^{1}_{\PP_{X}}$ follows from Assumption \ref{assconfarvarcon} and the integrability of $q$. Consequently, all the above results in this section remain valid even if the integrability assumption on $Y$ is relaxed to that of $q$.
\erem 
 

\brem
\label{rement}
For ${\varepsilon}\geq 0$ and $ \theP, \cH$ and $Z _{\nn}  $ as in Definition \ref{defradcom},  the ${\varepsilon}$  covering number of $\cH$ with respect to the empirical $L^{1}$  norm at $Z _{\nn}  $
is 
\begin{align*}
\cN_{1}(\cH,Z _{\nn}  ,{\varepsilon})\eqd \min\left\{m\in\N:  \exists \, g_{1:m} \in {(\cL( \theP))^{m}}: \sup_{h\in \cH}\min_{l\in 1\,..\, m}\frac 1n\sum_{k=1}^{n}|h(Z _{k})-g_{l}(Z _{k})|<{\varepsilon}\right\} 
\end{align*}
(with the convention $\inf\O=\infty$). 
 The quantity 
$
    \log(\cN_{1}(\cH,Z _{\nn}  ,{\varepsilon}))
$
is called the $L^1$ $\varepsilon$-entropy of $\cH$ at $Z_{\nn}$.
The interplay between the entropy and the Rademacher complexity is explained in Massart's Lemma \ref{lem:mas}, which illuminates the usefulness of having upper bounds on the entropy for applications of the bounds above.
\erem

\subsection{Estimation of ES-VaR: General Setting %
}
\label{secradconintes}

Note that two random variables whose cdf differ only beyond their (assumed common) VaR can have arbitrarily far away ES.  The truncated loss
$%
\lpesB (y,\theu,\thev)\eqdef
\Big(\Tb{
\lpvar_{\hun}(y,\theu  )-\theu
}-\thev \Big)^{2} 
$  
for some positive constant $B$ (compare with the second line in \eqref{equdefinslosvar})  is therefore introduced in view of establishing ES related concentrations.  
We assume $Y$  square integrable, so that $\cLabt\nnx\subset   L^{2}_{\PP_{X}} $,
by Lemma \ref{l:Linte}. We fix a hypothesis space $\cGB$ of $[0,B]$ valued  $\cB_{\cX}/\cB_{\R}$ measurable,  square integrable functions
  and, given $f\in\cLabt 
  $, we fix some function 
  \begin{align}
 \empappexpshoB{f}\in \argmin  _{g\in \cGB} \ave{n}{ \lpesB _{\hde}(Y_{i},f(X_{i}),g(X_{i}))},
\label{equappexpsho}
  \end{align} 
assumed to exist, interpreted as the best  empirical hypothesis for  $\ES-\VaR$ (after truncation by $B$) within $\cGB$, conditioned to the hypothesis $f$ for VaR. 

\subsection{A Priori Error Bounds for the Estimation of ES$-$VaR}
\label{ss:esv}

We are now in a position to establish our a priori bound on the error of the theoretical estimate $\empappexpshoB{f}$ with respect to $r$ given an hypothesis 
 $f$ for $q$.

\bt
\label{thebouerrrrad} In the  above setting, given $\delta\in (0,1)$, the inequality 
\beql{equesterrvardelentestes}
\|\empappexpshoB{f}-\ther\|_{\PP_{X},2}
\leq& \inf_{g\in \cGB}\|g-\ther\|_{\PP_{X},2}
\\
& +2\Big((1-\alpha)^{-1}\|f%
-\theq\|_{\PP_{X},2}+\|((1-\alpha)^{-1}( Y -\theq(X))^{+}-B)^+\|_{\PP,2}
\Big)
\\
& +  B\left(\frac{4\Radcave(\cGB,X_{\nn})}{B}   +\sqrt{\frac {2\log({2}/{\delta}) }{ n}}\right)^{1/2} 
\eeql
holds with probability at least $1-\delta$.
\et 
\begin{proof}
See Section \ref{secprolemristofun}.  \finproof
\end{proof}

\paragraph{Rademacher Versus Vapnik-Chervonenkis (VC) Regimes.}

An application of Massart's Lemma  \ref{lem:mas} 
 for 
$\varepsilon=B\sqrt{n}$ to 
\eqref{equesterrvardelentestes} yields the upper bound 
\beql{equesterrvardelentestes2}
&\|\empappexpshoB{f}-\ther\|_{\PP_{X},2}
\leq \inf_{g\in \cGB}\|g-\ther\|_{\PP_{X},2}
\\
&\qqq +{2}\Big((1-\alpha)^{-1}\|f%
-\theq\|_{\PP_{X},2}+\|((1-\alpha)^{-1}( Y -\theq(X))^{+}-B)^+\|_{\PP,2}
\Big)
\\
&\qqq + \frac {B}{{n}^{1/4}}\left(4\left(1+\Esp{\sqrt{2 \log (\cN_{1}(\cGB,X_{\nn}  ,B/\sqrt{n}) )}}\right)+
\sqrt{2\log({2}/{\delta})}
\right)^{1/2} 
\eeql
 with probability at least $1-\delta$.

According to well-known facts, both \eqref{equesterrvardelentestes} and \eqref{equesterrvardelentestes2} show a statistical error (the third line in both displays) with a suboptimal rate for the case at hand, namely $O(n^{-1/4})$ instead of $O(n^{-1/2})$.
An upper bound with ``the right rate'' $O(n^{-1/2})$ can be achieved via a separate Vapnik-Chervonenkis (VC) analysis that can be found in the arXiv v1 version of this work  
(based upon  \citet*[Theorem 2.2]{bargob19}, itself refining \citet*[Theorem 11.4 p.201]{gyor:kohl:krzy:walk:02}),
according to which the inequality
\beql{equinefas}
&\|\empappexpshoB{f}-\ther\|_{\PP_{X},2}\leq \sqrt{(6\,\lambda-5)} \inf_{g\in \cGB}\|g-
\ther\|_{\PP_{X},2}
\\
&\;+(1+\sqrt{(6\,\lambda-5)})((1-\alpha)^{-1}\|f%
-\theq\|_{\PP_{X},2}%
+ \|((1-\alpha)^{-1}( Y -\theq(X))^{+}-B)^+\|_{\PP,2} )
 \\
&\;+ \frac{2^3\, 3^{1/2}B}{\sqrt{(\lambda-1)n}}\Big( 
2\log\left(21\,\Esp{\cN_{1}\left(\cGB, X_{\nn}  ,B/(24n)\right)}\right)
+2\log\left(2/\delta\right)
\Big)^{1/2},
\eeql
holds with probability at least $1-\delta$ and for every $1< \lambda\leq 13/12$ (see also \citet[Remark 3.2]{bargob19}).

But the size of the constants involved in \eqref{equinefas} makes it less accurate than \eqref{equesterrvardelentestes2} in a relatively wide range of sample sizes. To give a crude comparison between the bounds \eqref{equesterrvardelentestes2} and \eqref{equinefas},
first note that, since $\sqrt{6\lambda-5}\approx 1$, it is reasonable to limit the discussion to a comparison between the terms in the third line of the inequalities (\ref{equesterrvardelentestes2}) and (\ref{equinefas}). If we focus on the dependence of the error bounds w.r.t. the data size $n$, and if we neglect the complexity terms by considering $\cN_1(\cdots)=1$, a consideration of the usual case $\delta=1/20$ and the choice $\lambda-1=1/12$ 
shows that the ratio between the two terms in the third line of the inequalities (\ref{equesterrvardelentestes2}) and (\ref{equinefas}) is (crudely) approximated by
\beql{e:ratio}
\frac{\left(\frac {{1}}{\sqrt{n}}\left({4}+
\sqrt{2\log({2}/{\delta})}
\right)\right)^{1/2}}{
2^{3}\, 3^{1/2}\Big( \frac{12}n\Big[2\log({21})+2\log\left({2}/{\delta}\right)
\Big]\Big)^{1/2}
}
&\stackrel{\delta=1/20}\approx
 {1.47\times 10^{-2}}\,n^{1/4}.
\eeql
This shows that the Rademacher-approach inequality \eqref{equesterrvardelentestes} is better than the VC-approach inequality \eqref{equinefas}, namely the right-hand side of \eqref{e:ratio} is smaller than $1$, whenever
 \begin{align}
 n \precapprox  2.14 \times 10^7\label{equdesineequ:thresold},
 \end{align}
 which can be understood as a  
 heuristic boundary between ``small-medium'' and ``big'' data, where we pass from the Rademacher to the VC regime.


\brem
\label{remboucom}
The bounds \eqref{equesterrvardelentestes} as well as \eqref{equesterrvardelentestes2} or \eqref{equinefas} above can be combined with the bound \eqref{equesterrvardel2} for giving rise to bounds on the estimate $\empappexpshoB{\empappvar}$ of $\ther$ based on the estimation $\empappvar$ of $\theq$ if $\empappvar, \theq \in \cLabt$, e.g. 
 \beql{e:fullbound}
 &\|\empappexpshoB{f}-\ther\|_{\PP_{X},2}
\leq \inf_{g\in \cGB}\|g-\ther\|_{\PP_{X},2} +2\|((1-\alpha)^{-1}( Y -\theq(X))^{+}-B)^+\|_{\PP,2}
\\
&\qq +\frac{2}{c(1-\alpha)^{3/2}}\Bigg[\left((2-\alpha)\inf_{f\in \cFab}\|f-\objfunvar\|_{\PP_{X},1}\right)^{1/2}\wedge 
 \left( C(1-\alpha)^{1/2}\inf_{f\in \cFab}\|(f-\objfunvar)\|_{\PP_{X},2}\right)
\\
&\qq\qqq +\left(2(2-\alpha)\Radcave(\cFab,X_{\nn}  )+ 
\|b-a\|_{\PP_X,\infty} \sqrt{\frac{2\log (2/\delta)}{n}}\right)^{1/2}\Bigg]\\
&\qq + \left(4B\Radcave(\cGB,X_{\nn}  ) +B^2\sqrt{\frac {2\log({2}/{\delta}) }{ n}}\right)^{1/2}.
 \eeql

\erem

\subsection{A Posteriori Monte Carlo Validation of VaR and ES learners}\label{ss:trick}
{\def\thisq{\check{q}}\def\thisq{f}
\def\this{\check{s}}\def\this{h}

The previous a priori error analysis is specific to the estimates 
$\empappvar$ and $\empappes{\empappvar} \eqd  \empappvar+  \empappexpsho{\empappvar}$ of $q$ and $s$. It is
 also theoretical in the sense that it supposes that global minimization is reached through training.
 
 By contrast, the following a posteriori error analysis can be applied to any tentative approximations $\thisq (X)$ of $q(X)=\VaR(Y|X)$ and $h(X)$ of $s(X)=\ES(Y|X)$ at the confidence level $\alpha$, including the ones that will arise from the practical training schemes of Sections
  \ref{sec:es_single_alpha} and 
\ref{par:var_cont_net}-\ref{par:var_interp_net},
 namely numerical stochastic gradient descent (SGD) approximations of $ \empappvar$ and $ \empappvar+ \empappexpsho{\empappvar}$. 
In fact, assuming one has access to the data generating process, as it is the case in most quantitative finance problems, one can  estimate distances of any guesses to the groundtruth (conditional) VaR and ES without directly computing  the latter, using an companion out-of-sample (dubbed ``twin'' in reference to $Y^{(1)}$ and $Y^{(2)}$ below) Monte Carlo procedure relying on the following result.

\bp\label{p:twin}  Let $Y^{(1)}$ and $Y^{(2)}$
denote two conditionally independent copies  of $Y$ given $X$, 
meaning that
for any bounded, $\cB_{\R}$ measurable functions $\rho$ and $\varrho$,  $$\Esp{\rho(Y^{(j)})|X}
=\Esp{\rho(Y)|X}\sp \Esp{\rho(Y^{(1)})\varrho(Y^{(2)})|X}=\Esp{\rho(Y^{(1)})|X}\Esp{\varrho(Y^{(2)})|X}.$$  
\noindent
{\rm \textbf{(i)}} For any ${\cB_{\cX}}/{\cB_\R}$  measurable function 
$\thisq$,  
\beql{eq:twinsim_var_err}
& \|\PPro{Y> \thisq (X)|X}-1+\alpha\|^2_{\PP ,2} = (1-\alpha) \times\\&\;\; \left(1-\alpha- \left( \PPro{Y^{(1)}>\thisq (X)}+  \PPro{Y^{(2)}>\thisq (X)}\right)\right) + \PPro{Y^{(1)}\wedge Y^{(2)}>\thisq (X)} .
 \eeql
 {\rm\hfill\break \textbf{(ii)}} For any $\thisq$ and $\this$  in $\cLabt$
under Assumption \ref{assconfarvarcon},
if $Y$ is square integrable and if, for some positive constant $c  $, $\inf_{y\in \big(a(x),b(x)\big)} \dot{F}_{Y|X=x}(y) \geq c$  holds $\PP_X$ a.s.,
then
\beql{eq:es_err_decompo}
 \|\this(X)-s(X) \|_{\PP ,2} &\le \left\|\this(X)-\thisq (X)-\Esp{{(1-\alpha)}^{-1}(Y-\thisq (X))^+|X} \right \|_{\PP ,2}\\
 &\qq+\frac{2-\alpha}{c(1-\alpha)}\|\PP [Y > \thisq (X)|X]-1+\alpha \|_{\PP ,2}, 
\eeql
where
\begin{align}
\begin{split}
&\left\|\this(X)-\thisq (X)-\Esp{{(1-\alpha)}^{-1}(Y-\thisq (X))^{+}|X}\right\|_{\PP ,2}^2 = \|\this(X)-\thisq (X)\|_{\PP ,2}^2\\
&\qqq+\frac{1}{(1-\alpha)^2}\Esp{(Y^{(1)}-\thisq (X))^{+}(Y^{(2)}-\thisq (X))^{+}}\\
&\qqq-\frac{2}{1-\alpha}\Esp{(\this(X)-\thisq (X))(Y-\thisq (X))^{+}}
\label{eq:twinsim_es_err_proxy}
\end{split}
\end{align}
and  $\frac{2-\alpha}{c(1-\alpha)}\|\PP [Y\geq \thisq (X)|X]-1+\alpha\|_{\PP ,2}$
is given by \eqref{eq:twinsim_var_err}.
\ep

\proof See Section \ref{ss:prooftwin}.  \finproof
\medskip


\noindent
The expectations and probabilities in \eqref{eq:twinsim_var_err} and \eqref{eq:twinsim_es_err_proxy} can be estimated by Monte Carlo simulation (see Algorithm \ref{alg:twin}), as opposed to nested Monte Carlo that would be required to explicitly attempt to approximate the conditional expectations involved in the left hand sides of
\eqref{eq:twinsim_var_err} and
\eqref{eq:twinsim_es_err_proxy}. Moreover the accuracy of the twin Monte-Carlo estimates can be controlled by computing confidence intervals.

As $1-\alpha=\PP [Y\geq q (X)|X]$ holds $\PP$ almost  surely,
the distance in \eqref{eq:twinsim_var_err} can be interpreted as a distance {in $p$-values}  between the quantile estimate
$f(X)$ and the true quantile $ {q}(X)$.
Note that because
\eqref{eq:es_err_decompo}
is only an inequality 
and due to the $\frac{1}{c}$ factor in \eqref{eq:twinsim_es_err_proxy}, the control 
on
$\|\this(X)-s(X)\|_{\PP ,2}$ provided by \eqref{eq:es_err_decompo} can be quite crude.
Hence  the control on
$\|\this(X)-s(X)\|_{\PP ,2}$ provided by twin Monte Carlo based on
\eqref{eq:twinsim_var_err}-\eqref{eq:es_err_decompo}-\eqref{eq:twinsim_es_err_proxy} 
can be quite conservative. In particular, such a procedure is appropriate for ensuring that 
$\|\this(X)-s(X)\|_{\PP ,2}$ is good enough for a given estimator $\this$ of $s$, but it cannot be used to compare two estimators of $s$. For comparing different estimators, however (of the expected shortfall or the quantile), twin Monte Carlo for e.g.\ $\Esp{(\this(X) - \mathrm{E}[\lpvar (Y, q (X))|X])^2}$ is not needed anyway, it is enough to compare the corresponding out-of-sample losses e.g.\ \bel&\Esp{(\this(X) - \lpvar (Y, q (X)))^2} = \Esp{(\this(X) - \mathrm{E}[\lpvar (Y, q (X))|X])^2} +\\&\qqq \Esp{(\lpvar (Y, q (X)) - \mathrm{E}[\lpvar (Y, q (X))|X])^2}. \eel

It may happen that the empirical version of  the right-hand side in \eqref{eq:twinsim_var_err}, or at least the lower bound of the corresponding Monte Carlo confidence interval, is negative, in which case the corresponding Monte Carlo estimate or lower bound cannot be used directly for the left-hand side. However, one could mitigate that by using a confidence upper-bound for the twin Monte-Carlo estimator, where a high enough confidence level can help get a more robust upper-bound for the a posteriori $L^2$ error.
Also, the fact that we chose  $\PPro{Y^{(1)}>\thisq (X)}+  \PPro{Y^{(2)}>\thisq (X)}$ rather than $2\PPro{Y^{(1)}>\thisq (X)}$ in \eqref{eq:twinsim_var_err} is arbitrary in this respect (and just motivated by the formal symmetry of the ensuing formula).

\begin{algorithm}[!t] 
 
\small
\LinesNumbered
\SetAlgoLined
\SetKwInOut{AlgName}{name}
\SetKwInOut{Input}{input}\SetKwInOut{Output}{output}
\AlgName{TwinVal}
\Input{out-of-sample $\{(X_i,Y_i^{(1)}, Y_i^{(2)})\}_{i = 1}^\thenu$ with $Y_i^{(1)}, Y_i^{(2)}$ independent copies of $Y$ given $X=X_i$, a confidence level $\alpha$, corresponding estimates $\thisq$ and $\this$ of $q$ and $s$, tolerance levels $\delta^{\rm var}$ and $\delta^{\rm es}$}
 
\Output{Quality of $\thisq$ and $\this$} 
Compute $(\epsilon^{\rm var})^2 = \frac{1}{\thenu}\sum_{i=1}^\thenu \big( (1-\alpha)(1-\alpha- \un_{Y^{(1)}_i>\thisq (X_i)}- \un_{Y^{(2)}_i>\thisq (X_i)}) + \un_{Y^{(1)}\wedge Y^{(2)}_i>\thisq (X_i)\big)} $ 

\eIf{$\epsilon^{\rm var}> \delta^{\rm var}$}{Reply already $\thisq$ is bad}{

Compute $( \epsilon^{\rm es})^2 = \frac{1}{\thenu}\sum_{i=1}^\thenu \Big[(\this(X_i)-\thisq (X_i))^2
+\frac{1}{(1-\alpha)^2}(Y_i^{(1)}-\thisq (X_i))^{+}(Y_i^{(2)}-\thisq (X_i))^{+}
-\frac{2}{1-\alpha}(\this(X_i)-\thisq (X_i))(Y_i^{(1)}-\thisq (X_i))^{+}]\Big] $

\eIf{$\epsilon^{\rm es}> \delta^{\rm es}$}{Reply $\thisq$ is good but $\this$ is bad}
{Reply $\thisq$ and $\this$ are good}

}

\caption{Twin Monte Carlo validation for VaR and ES.}
\label{alg:twin} 
\end{algorithm}

In the case where the twin Monte Carlo estimates for the right-hand-sides in \eqref{eq:twinsim_var_err}
and
\eqref{eq:twinsim_es_err_proxy}, after having been confirmed to be accurate by drawing enough samples, are not good enough, one can
{improve the numerical optimization, in first attempt, and then} act on the hypothesis space. 
For instance, in the case of the next sections of the paper where hypothesis spaces of neural networks are used,
one can improve the corresponding stochastic gradient descent by changing the optimizer  (e.g.~switching from the basic SGD of Algorithm \ref{alg:sgd} to
a more sophisticated Adam SGD as effectively done in our numerics),
in first attempt, and then try to train with more layers/units or better architectures.

\section{Learning Using Neural Networks\label{sec:learningwithneuralnets}}
\def\tx{\tilde{x}}
In this section we follow up on the theory of Section \ref{algo:ca} in the case of learning with the following class of neural networks that will be used as hypothesis spaces.
\bd\label{d:network}
 Let $\sigma$ be a 1-Lipschitz, positive-homogeneous activation function (such as ReLU), applied element-wise when supplied with a vector as input, and let $(d,o,l, m,B_{1:(l+1)})\in \N^4 \times (0,\infty)^{l+1}$. We consider the family of neural networks on $\cX \subseteq\R^{d} $ with $o$ outputs, $l$ hidden layers, $m$ (or less)
 hidden units, activation function $\sigma$, and Euclidean norm regularisation on the weights and bias, i.e., with $\tx\eqdef \begin{bmatrix} x \\ 1 \end{bmatrix}$ introduced to take into account the bias,
 \beql{e:defnn}
&\cNN(d,o,B_{1:(l+1)},l,m,\sigma)\eqd \Big\{\R^d\ni x  
 \mapsto W_{l+1}\sigma(W_{l}\sigma(\dots\sigma(W_1 \tx)) \in\R^o\, ; \\
 &\qqq W_1\in \R^{m\times (d+1)}, W_{2:l}\in (\R^{m\times m})^{l-1}, W_{l+1} \in \R^{o \times m}, |W_k|_2\leq B_k \mbox{ for }k=1\dots l+1
 \Big\}.
 \eeql
\ed
 The norm $|\,\cdot\,|_2$, defined as the square root of the sum of the squares of all entries of  $\cdot$\,, is called the Frobenius norm when $\cdot$ is a matrix.  It is not difficult to see that, for matrices (or vectors) $W$ and $V$ such that the product $WV$ is well defined,
 \beql{mulprofro}
     |WV|_{2}\leq |W|_{2}|V|_{2}.
 \eeql

\subsection{Error Bound of 
the Single-$\alpha$ Learning
Algorithm 
 With Neural Networks\label{ss:nnan}}

\bl \label{l:radenn}
For any $f\in \cNN(d,1,B_{1:(l+1)},l,m,\sigma)$, we have 
\beql{e:bound_nn}
|f(X)| \leq (|X|_2+1)
\prod_{k=1}^{l+1}B_k .
\eeql
If  
$\cX$ is 
compact , then $\cNN(d,1,B_{1:(l+1)},l,m,\sigma)$ is uniformly bounded and
\beql{e:rade_nn_relu}
\Radcave(\cNN(d,1,B_{1:(l+1)},l,m,\sigma),X_{\nn}  ) \leq 
\frac{\sqrt{2(l+1)\log 2}+1}{\sqrt{n}} (\||X|_2\|_{\PP_X, \infty}+1)\prod_{k=1}^{l+1}B_k.
 \eeql 
 \el
 \proof See
 Section \ref{s:bnn}.  \finproof\\
 
 The estimate \eqref{e:rade_nn_relu} can be combined with  Theorem \ref{theeststaerrvar} to provide a more explicit error control on $\empappvar$.

 \bt\label{t:cvrateNNq}
In the setup of Theorem \ref{theeststaerrvar} with $-a(X) =b(X) = (|X|_2+1) \prod_{k=1}^{l+1}B_k$ and $\cFab = \cNN(d,1,B_{1:(l+1)},l,m,\sigma)$, 
assume that
 \begin{itemize}
 \item[(i)] $\cX$ is a compact subset of $\R^d$,
     \item[(ii)]  $ c \leq \gamma_f(X) \leq C$ holds $\PP$  a.s. for some constants $0<c\leq C<\infty$ independent of $f\in \cFab$.
 \end{itemize}
Then 
there exist $\appvar$ and  $\empappvar$ satisfying \eqref{equappvarprel}. For any $\delta\in (0,1)$,
\beql{equesterrvardelentestneunet}
&{c}(1-\alpha)^{1/2}\|\empappvar-\theq\|_{_{\PP_{X},2}}
 \leq \left((2-\alpha)\inf_{f\in \F}\|f-\objfunvar\|_{\PP_{X},1}\right)^{1/2}\wedge \left(C(1-\alpha)^{1/2}\inf_{f\in \cFab}\|f-\objfunvar\|_{\PP_{X},2}\right)
 \\ 
&\;+ 
\sqrt{2} \left(\frac{(2-\alpha)(\sqrt{2(l+1)\log 2}+1)}{\sqrt{n}} + \sqrt{\frac{2\log (2/\delta)} {n}} \right)^{1/2}\left((\||X|_2\|_{\PP_X,\infty}+1)\prod_{k=1}^{l+1}B_k\right)^{1/2}
\eeql
holds  with probability at least $1-\delta$. 
\fe
 \et

\proof  See Section \ref{s:proof_nnvar}.  \finproof

\brem Under the regularization embedded in
$\cFab=\cNN(d,o,B_{1:(l+1)},l,m,\sigma)$,
the estimation
error term (second line) in \eqref{equesterrvardelentestneunet} does not
depend on the number of neurons $m$ nor on the dimension $d$ of the input space.  
The bias term in the first line is decreasing in $m$ (via $\cFab$) and potentially increasing in $d$ (via $\cFab$ and $q$).
\erem

\brem\label{r:rate} Our convergence rate $n^{-1/4}$ for the satistical error is consistent with \citet[Theorem 2]{padilla2020quantile}, which however only deals with the case of deterministic $X_i$ and the corresponding empirical measure. \citeN{shen2021deep} requires integrability of $Y$ of order $p>1$, with
rates better than $n^{-1/4}$ for $p>2$. On the other side, both papers rely on VC arguments leading to unspecified constants in their statistical error bounds, see also our discussion below Theorem \ref{thebouerrrrad}.
\erem

An analogous reasoning, using this time Theorem \ref{thebouerrrrad}, leads to the following error control on $\empappexpsho{f%
}$ (with $(\cH)^+=\left\{h^+: h\in\cH\right\}$ for any set of functions $\cH$):
 
 \bt\label{t:cvesnn}
In the setup  of Theorem \ref{thebouerrrrad} with $\cX$ compact, $B = (\||X|_2\|_{\PP_X, \infty}+1) \prod_{k=1}^{l+1}B_k $, 
and $\cGB= (\cNN(d,1,B_{1:(l+1)},l,m,\sigma))^+$, there exists a function  $\empappexpshoB{f}$ satisfying \eqref{equappexpsho}. For any $\delta\in (0,1)$ and $f\in\cLabt$,
\beql{equesterresdelentestneunet}
&\|\empappexpsho{f%
}-\ther\|_{\PP_{X},2}
\leq \inf_{g\in \cGB}\|g-\ther\|_{\PP_{X},2}
\\
&\qq+ 2 \left((1-\alpha)^{-1}\|f%
-\theq\|_{\PP_{X},2}+\left\|\left(\frac{( Y -\theq(X))^{+}}{1-\alpha}-(\||X|_2\|_{\PP_X, \infty}+1) \prod_{k=1}^{l+1}B_k\right)^{+}\right\|_{\PP,2}
\right)
\\
&\qq+    \left( \frac {4(\sqrt{2(l+1)\log 2}+1)}{\sqrt{n}} +\sqrt{\frac {2\log({2}/{\delta})} {n}}\right)^{1/2} (\||X|_2\|_{\PP_X, \infty}+1)\prod_{k=1}^{l+1}B_k 
\eeql
 holds  with probability at least $1-\delta$.\fe
 \et

 \proof See Section \ref{s:proof_esnn}.  \finproof

\subsection{Algorithms\label{sec:es_single_alpha}}

 \begin{algorithm} 
\small
\LinesNumbered
\SetAlgoLined
\SetKwInOut{AlgName}{name}
\SetKwInOut{Input}{input}\SetKwInOut{Output}{output}
\AlgName{SGDOpt}
\Input{$(X, Y)_{\nn}  $, a partition $\Batches$ of $\{1 \dots \then \} $, a number of epochs $E\in\mathbb{N}^{\star}$, a learning rate $\eta>0$, initial weight (matrix) $\widehat{W}$ parameters,  a loss function $\pin =\pin  (W, \text{batch})$ and a regularisation weight $\kappa$ (set at 0.01 by default)}

\Output{Trained parameters $\widehat{W}$} 
Set $\pin  (W, \text{batch}) = \pin  (W, \text{batch}) + \kappa |W|_2$

 \For(\tcp*[f]{loop over epochs}){$\text{epoch} = 1, \dots, E$}{
  \For(\tcp*[f]{loop over batches}){$\text{batch} \in \Batches$}{
    $\widehat{W} \leftarrow \widehat{W} - \eta \nabla_{W}\pin (\widehat{W}, \text{batch})$ \\ 
  }
 }
\caption{Mini-batch stochastic gradient descent in a neural net hypothesis space.} 
\label{alg:sgd}
\end{algorithm}
In practice the hard constraints $B_{1:(l+1)}$ of the neural network family are ``softly'' handled by penalization: see Algorithm \ref{alg:sgd}. Accordingly, we drop $B_{1:(l+1)}$ and simplify the notation for the network family to $\cNN(d,o,l,m,\sigma)$ in what follows. Let $\zeta^{d, o}_{\lay +1}(x, W)$ denote  a function in the family $\cNN(d,o,l,m,\sigma)$ with the Softplus activation function, i.e.\ $\sigma(x) = \log(1+\exp(x))$, where $W$ represents the set of network parameters. We observed numerically similar performances between ReLU and Softplus networks.  Despite  our theoretical bounds in Section \ref{ss:nnan} being built upon the ReLU network, we use the Softplus networks in our numerics below for their ability to provide analytical derivatives of their outputs with respect to their inputs, as required in Section \ref{sec:multi_alpha}. 

Given an i.i.d sample $(X, Y)_{\nn}  $ of $(X, Y)$ as before,
instead of considering global argminima $\empappvar$ and (given $\empappvar$) $  \empappexpsho{\empappvar}$ 
of the related empirical losses as in the theoretical analysis of Sections \ref{ss:vars}--\ref{ss:esv}, we consider various mini-batch  SGD estimates  (cf.\
Algorithm \ref{alg:sgd}) for 
$\empappvar$ and $\empappes{\empappvar}$ . 
These SGD estimates are then amenable to the a posteriori (twin Monte Carlo) error analysis
of Section \ref{ss:trick}.

\begin{algorithm}
\small
\LinesNumbered
\SetAlgoLined
\SetKwInOut{AlgName}{name}
\SetKwInOut{Input}{input}\SetKwInOut{Output}{output}
\AlgName{VaRAlg}
\Input{$(X, Y)_{\nn}  $, a partition $\Batches$ of $\{1 \dots \then \} $, a quantile level $\alpha$, a number of epochs $E\in\mathbb{N}^{\star}$, a learning rate $\eta>0$, initial values for the network parameters $\widehat{W}$, and neural network output function $\zeta^{d, 1}_{\lay +1}(x, W)$}

\Output{Trained parameters of VaR network} 
define $\displaystyle\pin^{\rm var}(W , \text{batch})= \frac{1}{|\text{batch}|}\sum_{i\in \text{batch}} [(Y_i-\zeta^{d, 1}_{\lay +1}(X_i, W))^++(1-\alpha)\zeta^{d, 1}_{\lay +1}(X_i, W)]$\\
$\widehat{W}^{\rm var} \leftarrow  \text{SGDOpt}((X, Y)_{\nn}  , \Batches, E, \eta, \widehat{W}, \pin^{\rm var} 
)
$
\caption{Neural network regression for learning the VaR.} 
\label{alg:var}
\end{algorithm}

Algorithm \ref{alg:var} thus produces a VaR predictor
$$
\zeta^{d, 1}_{\lay +1}(X, \widehat{W}^{\rm var})\approx \empappvar(X) 
$$
in the form a function of $X$ represented by a neural network from  $\cNN(d,1,l,m,\sigma)$,
for given $m$ and $\lay$. 

\begin{algorithm}
\small
\LinesNumbered
\SetAlgoLined
\SetKwInOut{AlgName}{name}
\SetKwInOut{Input}{input}\SetKwInOut{Output}{output}
\AlgName{ESAlg}
\Input{$(X,Y)_{1:\then}$, a partition $\Batches$ of $\{1 \dots \then \} $, a quantile level $\alpha$, a number of epochs $E\in\mathbb{N}^{\star}$, a learning rate $\eta>0$, initial values for the network parameters $\widehat{W} $ and neural network output function $\zeta^{d, 1}_{\lay +1}(x, W)$}
 
\Output{Trained parameters of ES network $\widehat{W}^{\rm es}$} 

\tcp{Learn the corresponding VaR}
$\widehat{W}^{\rm var} \leftarrow \text{VaRAlg}((X,Y)_{1:\then}, \Batches, \alpha, E, \eta,\widehat{W} )$ \\

\eIf{ linear regression}{
\tcp{Remind $W_k$ denote the weight of $k$  th layer}
\beql{bluf}
&\{\widehat{W}^{\rm es}_k\}_{k=1}^l  \leftarrow \{\widehat{W}_k^{\rm var}\}_{k=1}^l \\
&\widehat{W}^{\rm es}_{l+1} \leftarrow   \argmin_{W_{l+1}}
\frac{1}{n} \sum_{i=1}^n \Big[
{(1-\alpha)}^{-1}\Big(Y_i-\zeta^{d, 1}_{\lay +1}(X_i, \widehat{W}^{\rm var})\Big)^+ \\
    &\quad+\zeta^{d, 1}_{\lay +1}(X_i, \widehat{W}^{\rm var}) -\zeta^{d, 1}_{\lay +1} (X_i, (\{\widehat{W}^{\rm var}_k\}_{k=1}^l,W_{l+1}))\Big]^2
\eeql 
}{
define 
$\pin^{\rm es}(W , b, \text{batch})= \frac{1}{|\text{batch}|}\sum_{i\in \text{batch}} [( {(1-\alpha)}^{-1}(Y_i-\zeta^{d, 1}_{\lay +1}(X_i, \widehat{W}^{\rm var}))^+  +\zeta^{d, 1}_{\lay +1}(X_i, \widehat{W}^{\rm var}) -\zeta^{d, 1}_{\lay +1}(X_i, W))^2]
$

$\widehat{W}^{\rm es}  \leftarrow  \text{SGDOpt}((X, Y)_{\nn}  , \Batches, E, \eta, \widehat{W}^{\rm var}, \pin^{\rm es})$\\
}

\caption{Neural network regressions for learning the ES in two steps, using least-squares regression of the linear readout map  or full NN training or for deducing the ES from the VaR.}

\label{alg:es}
\end{algorithm}
Given this 
VaR predictor,
Algorithm \ref{alg:es} produces
$$
\zeta^{d, 1}_{\lay +1}(X, \widehat{W}^{\rm es})\approx \empappes{\empappvar}(X) ,
$$
in two possible ways: either by training a dedicated neural network from scratch, or  by using the same 
architecture as the one used for the VaR, freezing the weights of all hidden layers as those of the VaR network and 
least-square regressing the linear-readout map.
We show in Section \ref{sec:toymodel} that such a transfer learning scheme is enough to obtain good approximations,
while also being very fast (a fraction of a second in our experiments) using highly optimized linear algebra routines such as the ones implemented by cuBLAS for Nvidia GPUs.

\section{Multi-$\alpha$ Learning for VaR\label{sec:multi_alpha}}

In this part we are interested in learning $\VaR(Y|X)$ for multiple confidence levels $\alpha\in(0,1)$ using a single empirical error minimization. This can help give insights into the sensitivity of $\VaR(Y|X)$ with respect to the confidence level, or into the full distribution of $Y$ given $X$ (e.g.\
approximated by a histogram representation).
 Given that the previous trainings were done for a single fixed confidence level $\alpha$,
we refer to them as {the single-$\alpha$ learning} (or {single-$\alpha$} for brevity in the numerics): under this approach, if one is interested in finding the conditional VaR for another confidence level, one has to repeat the training procedure using the new confidence level.

By contrast, the
{multi-$\alpha$ learning} approaches below allow learning $\VaR(Y|X)$ for multiple confidence levels within a single simulation run.
Regarding the  learning of $\ES(Y|X)$ for multiple confidence levels, the transfer learning trick (cf.\ the linear regression case in Algorithm \ref{alg:es}) in Section \ref{sec:es_single_alpha} was found to provide the most valuable alternative, whether  done $\alpha$ by $\alpha$ (our choice in the numerics below), as each run of it is very fast, or globally across $\alpha$'s based on either of the multi-$\alpha$ VaR approaches below. Hence we focus on the multi-$\alpha$ learning of VaR in what follows.

\subsection{The Crossing Quantile Issue\label{ss:qci}}
 When several quantile levels $\alpha$ are considered, a flaw inherent to linear quantile regression  
is the problem of crossing quantile curves, i.e.~the violation of the monotonicity with respect to $\alpha$.
The simultaneous learning of conditional quantiles for multiple confidence levels and the problem of quantile crossing were early addressed in \citeN{he1997quantile}, \citeN{koenker2004quantile} and \citeN{takeuchi2006nonparametric}, see also \citeN{moon2021learning} for a review of more recent references. 
To deal with the quantile crossing problem, two strategies for constraints can be considered. 

The first strategy is to consider explicitly the non-crossing constraints during the learning phase of the model in form of either hard constraints (that the model must strictly satisfy) or soft constraints (i.e.\ penalization). Once the non-crossing hard constraints are employed, the model is usually learned using primal-dual optimization algorithms. The latter are applicable in a wide class of models, e.g.~support vector regression \citep*{takeuchi2006nonparametric, sangnier2016joint} and spline regression \citep*{bonetal2010}, but notably not in the case of the family of (deep) neural networks, because of the computational cost and the poor scalability of projected gradient descent. Therefore, the non-crossing constraints are more preferably embedded in the training of neural networks via a penalty term, based in \cite{moon2021learning}
on a finite difference of the output of the neural network (that approximates the value-at-risk) for two confidence levels.

In Section \ref{par:var_cont_net} we use a similar penalization strategy, where, instead of penalizing the negative part of a finite difference, we penalize the negative part of the partial derivative of the network with respect to the confidence level. The partial derivative gives more information about the local behavior around training points and we can penalize its negative part at every $\alpha$ that appears at the training stage, e.g.\ for several thousands values of $\alpha$ in our numerics below, as opposed to penalizing negative increments at a few fixed values of $\alpha$ as in  \citet{moon2021learning}. 
Our approach also spares one hyperparameter, namely the size of the discrete increment in confidence levels used for the finite differences.

The second strategy is to use hypothesis spaces of functions nondecreasing with respect to the confidence level. \citeN{meinshausen2006quantile} introduce quantile regression forests. In this model the predicted quantile of a new point is based on the empirical percentile of the group (i.e.~the terminal leaf of each tree) where this point belongs, hence the monotonicity of the quantile estimates is satisfied by construction. Regarding neural networks, 
\citeN{hatalis2017smooth} propose a specific initialization scheme for the weights of the output layer, which does not prevent quantile crossings, but appears to reduce them significantly in their experiments. \citeN{cannon2018non} considers the confidence level as an additional explanatory variable and then explores a network such that the estimate is monotone with a defined covariate (confidence level), imposing the non-crossing. 
\citeN{gasthaus2019probabilistic} and \citet*{padilla2020quantile} use a (deep) network with multiple outputs, constrained by design to be positive, which are expected to approximate quantile increments. 

The latter resembles our 
approach in Section \ref{par:var_interp_net}{,
but} we sample the confidence level uniformly on a given interval and we further interpolate linearly with respect to the confidence level before insertion of the output of the neural network in the training loss (cf.~\eqref{num_var_multi_alpha_interpolation_finite_sample}-\eqref{num_var_multi_alpha_interpolation_network}),
in order to have a conditional quantile function that is valid for all quantile levels in the interval. %

\subsection{Extension of the Bounds to Multi-{$\alpha$} Learning}
\begin{table}[bp!]
\def\arraystretch{1.6}
\small
\centering
\begin{threeparttable}
\resizebox{\textwidth}{!}{
\begin{tabular}{|l|l|} 
\hline
{\bfseries Single-$\alpha$} & {\bfseries Multi-$\alpha$}  \\
\hline
 $X   $ valued in $\cX $   &  $(\alpha, X )   $  valued in $ \cI\otimes\cX $  \\  
\hline
$ F_{Y|X}(a(X))< \alpha < F_{Y|X}(b(X)), \; \PP\mbox{-a.s.}$ & $ F_{Y|X}(a(X))< \underline{\alpha} \leq \overline{\alpha} < F_{Y|X}(b(X)), \; \PP\mbox{-a.s.}$
\\ 
\hline
$\lpvar(y,u)={(1-\alpha)}^{-1}(y-\theu  )^++u$ & $\lpvar(\alpha, y,u)={(1-\alpha)}^{-1}(y-\theu  )^++u$    \\ 
\hline
$\lmeavar(f)= \Esp{\lpvar_{\hun}\left(\Yab ,f(X )\right)  \big| (X,Y)_{\nn}   }$ & $\lmeavar(f)= \Esp{\lpvar_{\hun}\left(\alpha,\Yab ,f(\alpha, X )\right)  \big| (\alpha, X,Y)_{\nn}   } $    \\ 
\hline
spaces $\cFab\subseteq\cLab 
$ of functions on $\cX$
& spaces $\cFab\subseteq\cLab
$  of functions on $\cI\otimes\cX$  \\ 
\hline
$\tilde{q}\in\argmin_{f\in\cFab} \lmeavar(f)$ & $\tilde{q}\in\argmin_{f\in\cFab} \lmeavar(f)$    \\ 
\hline
$\empappvar\in\argmin_{f\in\cFab}\frac{1}{n}\sum_{i =1}^n\lpvar(\Yab_i, f(X_i))$ & $\empappvar\in\argmin_{f\in\cFab}\frac{1}{n}\sum_{i =1}^n\lpvar(\alpha_i, \Yab_i, f( \alpha_i, X_i))$    \\ 
\hline
$\gamma_{f}(x)\eqd \Gamma_{{F_{Y|X=x}}}(f(x),\theq(x))$\tnote{$*$} &
$\gamma_{f}(\alpha,x)\eqd \Gamma_{{F_{Y|X=x}}}(f(\alpha,x),\theq(\alpha,x))$\tnote{$*$} \\ \hline
$\begin{aligned}
&\lmeavar(\appvar)-\lmeavar(\theq)=\|{\gamma_{\appvar}}(\appvar-\objfunvar)\|_{\PP_{X},2}^{2} = \\
&\;\inf_{f\in \cFab}\|{\gamma_{f}}(f-\objfunvar)\|_{\PP_{X},2}^{2}\leq \frac{2-\alpha}{1-\alpha}\inf_{f\in \cFab}\|f-\objfunvar\|_{\PP_{X},1}
\end{aligned}$ 
& 
$\begin{aligned}
&\lmeavar(\appvar)-\lmeavar(\theq)=\|{\gamma_{\appvar}}(\appvar-\objfunvar)\|_{\PP_{\alpha,X},2}^{2} = \\
&\;\inf_{f\in \cFab}\|{\gamma_{f}}(f-\objfunvar)\|_{\PP_{\alpha, X},2}^{2}\leq \frac{2-\overline{\alpha}}{1-\overline{\alpha}}\inf_{f\in \cFab}\|f-\objfunvar\|_{\PP_{\alpha, X},1}
\end{aligned}$  \\ \hline
$\begin{aligned}
    & (1-\alpha)^{-1}\Big(2(2-\alpha)\Radcave(\cFab,X_{\nn}  )\\
    &\qqq\qqq+ \frac{\|b-a\|_{\PP_X,\infty}\sqrt{2\log (2/\delta)}}{\sqrt{n}}\Big)
\end{aligned}$ & 
$\begin{aligned}
    &(1-\overline\alpha)^{-1}\Big(2(2-\overline\alpha)\Radcave(\cFab,(\alpha,X)_{\nn}  )\\
    &\qqq\qqq+ \frac{\|b-a\|_{\PP_X,\infty}\sqrt{2\log (2/\delta)}}{\sqrt{n}}\Big)
\end{aligned}$
\\\hline
$\begin{aligned}
    &\Radcave(\cNN(d,1,B_{1:(l+1)},l,m,\sigma),X_{\nn}  ) \\
    &\leq
\frac{\sqrt{2(l+1)\log 2}+1}{\sqrt{n}}(\||X|_2\|_{\PP_X, \infty}+1)\prod_{k=1}^{l+1}B_k 
\end{aligned}$ & 
$\begin{aligned}
    &\Radcave(\cNN(d+1,1,B_{1:(l+1)},l,m,\sigma),(\alpha, X)_{\nn}  ) \\
    &\leq
\frac{\sqrt{2(l+1)\log 2}+1}{\sqrt{n}}(\overline\alpha+\||X|_2\|_{\PP_X, \infty}+1) \prod_{k=1}^{l+1}B_k
\end{aligned}$
\\
\hline \hline
$\begin{aligned}
&es(u) = (1-\alpha)^{-1}\int_{\theu}^{\infty}( y - \theu )^{+}F(dy)+\theu,\\
&\Delta_{F}(u,\theq)\eqd  es(u)-es(q) \\
&\Gamma_{F}(\theu,\theq)\eqd \frac{\Delta_{F}(\theu,\theq)}{(\theu-\theq)^{2}}\textbf{1}_{(0,\infty)}(|\theu-\theq|)\\
&\qqq\qq+\frac{\den (\theq)}{2(1-\alpha)}\textbf{1}_{\{0\}}(\theu-\theq)
\end{aligned}$ & 
$\begin{aligned}
&es(\alpha, u) = (1-\alpha)^{-1}\int_{\theu}^{\infty}( y - \theu )^{+}F(dy)+\theu,\\
&\Delta_{F}(\alpha, u,\theq(\alpha))\eqd  es(\alpha, u)-es(\alpha, q(\alpha)) \\
&\Gamma_{F}(\alpha, \theu,\theq(\alpha))\eqd \frac{\Delta_{F}(\alpha,\theu,\theq(\alpha))}{(\theu-\theq(\alpha))^{2}}\textbf{1}_{(0,\infty)}(|\theu-\theq(\alpha)|)\\
&\qqq\qqq+\frac{\den (\theq(\alpha))}{2(1-\alpha)}\textbf{1}_{\{0\}}(\theu-\theq(\alpha))
\end{aligned}$ \\
\hline
\end{tabular}}
\caption{Main changes required to adapt the previous results and proofs from a single-quantile to a multi-quantile regression setup. }\label{e:singvsmul}
\begin{tablenotes}
\item[$*$] defined from the unconditional notation displayed after the double line.
\end{tablenotes}
\end{threeparttable}
\end{table}
The various proofs and bounds presented in this paper for a fixed $\alpha\in[0,1]$ can be extended to the multi-$\alpha$ learning framework where $\alpha$ is now a random variable supported by $ \cI=[\underline{\alpha}, \overline{\alpha}]\subseteq (0,1)$, treated as a new covariate alongside $X$. Hereafter we randomize $\alpha$ assuming $\alpha\sim\cU([\underline{\alpha}, \overline{\alpha}])$. We then consider a finite i.i.d sample $\alpha_1, \dots, \alpha_n$ of $\alpha$, independent of  covariates $(\alpha,X)$ and of the sample $(X,Y)_{1:n}$, and the loss functions that appear, together with  corresponding changes to the results above, in Table \ref{e:singvsmul}. 
The implementation of this approach using neural networks is discussed below.

\subsection{Learning With a Continuum of $ \alpha$\label{par:var_cont_net}}

The finite-sample training problem for this approach can be stated as follows:
\begin{equation}
 \argmin  _{W} \frac{1}{\then}\sum_{i=1}^\then[(Y_i-\zeta^{d+1, 1}_{\lay +1}([\alpha_i, X_i], W))^++(1-\alpha_i)\zeta^{d+1, 1}_{\lay +1}([\alpha_i, X_i], W)],
\label{num_var_multi_alpha_continuous_finite_sample}
\end{equation}
where $[a, x]$ denotes the vector obtained by concatenating a vector $x$ to a real $a$.
 One can also approximately impose the non-crossing of the quantiles by penalizing the sample average of the negative part of the partial derivative $ \partial_\alpha \zeta^{d+1, 1}_{\lay +1}([\alpha, X], W)$, as per
\begin{dmath}
 \argmin_{W} \frac{1}{\then}\sum_{i=1}^\then\Big[(Y_i-\zeta^{d+1, 1}_{\lay +1}([\alpha_i, X_i], W))^+ +(1-\alpha_i)\zeta^{d+1, 1}_{\lay +1}([\alpha_i, X_i], W)+\lambda\Big( \partial_\alpha \zeta^{d+1, 1}_{\lay +1}([\alpha_i, X_i], W)\Big)^-\Big],
\label{num_var_multi_alpha_continuous_penalized_finite_sample}
\end{dmath}
where $\lambda>0$ determines the strength of the penalization. 
\begin{algorithm}
\small
\LinesNumbered
\SetAlgoLined
\SetKwInOut{AlgName}{name}
\SetKwInOut{Input}{input}\SetKwInOut{Output}{output}
\AlgName{MultiContinousVaRAlg}
\Input{$(X,Y)_{ \nn    }$, a partition $\Batches$ of $\{1 \dots \then \} $, a quantile upper bound level $\overline{\alpha}$, and lower bound level $\underline{\alpha}$, a number of epochs $E\in\mathbb{N}^{\star}$, a learning rate $\eta>0$, a regularisation parameter $\lambda \geq 0$, initial values for the network parameters $\widehat{W} $ and neural network output function $\zeta^{d+1, 1}_{\lay +1}([a, x], W)$}
 
\Output{Trained parameters of multi-VaR network $\widehat{W}$} 

\tcp{Sample quantile levels $\alpha$}
$\alpha_i \sim \text{Uniform}(\underline{\alpha}, \overline{\alpha}) \quad \text{for } i = 1\dots \then$ \\
\tcp{Define a loss function}
\eIf{ non-crossing quantile regularisation}{\tcp{multi-$\alpha$(I)}

define $\displaystyle\pin^{\rm vars} (W , \text{batch})=\frac{1}{|\text{batch}|}\sum_{i\in \text{batch}} [(Y_i-\zeta^{d+1, 1}_{\lay +1}([\alpha_i, X_i], W))^++(1-\alpha_i)\zeta^{d+1, 1}_{\lay +1}([\alpha_i, X_i], W)+\lambda(\partial_\alpha \zeta^{d+1, 1}_{\lay +1}([\alpha_i, X_i], W))^-]$ \\
where $\partial_\alpha \zeta^{d+1, 1}_{\lay +1}([\alpha_i, X_i], W))^-]$ can be quickly computed

}{\tcp{multi-$\alpha$(II)}

define $\displaystyle\pin^{\rm vars} (W , \text{batch})=\frac{1}{|\text{batch}|}\sum_{i\in \text{batch}} [(Y_i-\zeta^{d+1, 1}_{\lay +1}([\alpha_i, X_i], W))^++(1-\alpha_i)\zeta^{d+1, 1}_{\lay +1}([\alpha_i, X_i], W)]$}

$\widehat{W}^{\rm vars}  \leftarrow  \text{SGDOpt}((X,Y)_{1:\then}, \Batches, E, \eta, \widehat{W},\displaystyle\pin^{\rm vars} )$
\caption{Learning multi continuous VaR.} 
\label{alg:vars}
\end{algorithm}
Algorithm \ref{alg:vars} thus produces
$$
\zeta^{d+1, 1}_{\lay +1}(\alpha,X, \widehat{W}^{\rm vars})\approx \widehat{q}
_{\alpha} (X) ,
$$
 where $\empappvar_{\alpha}(X)$ denotes the theoretical estimator (multi-$\alpha$ analog of $\empappvar (X)$ before, ignoring the SGD numerical optimization error) of the  $\VaR(Y|X)$ quantile function at any quantile level in $[\underline{\alpha}, \overline{\alpha}]$. Note that for non-zero $\lambda$, one uses a penalized pinball loss function that is beyond the scope of Theorems \ref{theeststaerrvar}-\ref{t:cvrateNNq}. The a posteriori Monte Carlo validation technique of Section \ref{ss:trick}, though, is still applicable to this (as to any) estimator.
Importantly, one can compute the derivative in \eqref{num_var_multi_alpha_continuous_penalized_finite_sample} fast in closed-form given our neural network parametrization, as 
$\partial_\alpha \zeta^{d+1, 1}_{\lay +1}([\alpha, X], W)  = W_{\lay +1}\partial_\alpha \zeta^{d+1, 1}_{n}([\alpha, X], W)$, where 
\begin{align*}
\partial_\alpha \zeta^{d+1, 1}_{0}([\alpha, X], W) &= [1, 0_d] \mbox{ and, for } k=1, \ldots,\lay,\\
\partial_\alpha \zeta^{d+1, 1}_{k}([\alpha, X], W) &= (W_k\partial_\alpha \zeta^{d+1, 1}_{k-1}([\alpha, X], W))\odot\sigma'(W_k\zeta^{d+1, 1}_{k-1}([\alpha, X], W)).
\end{align*}
Here $\odot$ is an element-wise product and $\sigma'$ is the derivative of $\sigma$ (applied element-wise). Given the computations of  $\zeta^{d+1, 1}_{\lay +1}([\alpha, X], W)$ and $\partial_\alpha \zeta^{d+1, 1}_{\lay +1}([\alpha, X], W)$ share many common sub-expressions, the recursions can be done at the same time, i.e.~at each $k\in\{0, \dots, \lay +1\}$, compute $\zeta^{d+1, 1}_k([\alpha, X], W)$ and then reuse the common sub-expressions to compute also $\partial_\alpha \zeta^{d+1, 1}_k([\alpha, X], W)$. In the numerics, we refer to this approach with 
{multi-$\alpha$(I)} if we use a non-zero $\lambda$, and {multi-$\alpha$(II)}
otherwise.

\subsection{Learning Via a Discrete Set of $\alpha$'s and Linear Interpolation\label{par:var_interp_net}} 

Another approach for multi-$\alpha$ learning is to use a finite set of confidence levels $\alpha^{(1)}< \dots<\alpha^{(o)}$ in $[\underline{\alpha}, \overline{\alpha}]$ in conjunction with linear interpolation. More precisely, we solve 
\beql{num_var_multi_alpha_interpolation_finite_sample}
&
\argmin  _{W} \frac{1}{\then}\sum_{i=1}^\then\Big[\Big(Y_i-\Sigma\big(\alpha_i,  \zeta^{d, o}_{\lay +1}( X_i, W)\big)\Big)^++(1-\alpha_i)\Sigma\big(\zeta^{d, o}_{\lay +1}( \alpha_i , X_i, W)  \big)\Big] ,
\eeql
where,  for $y=(y_0,\dots,y_{o-1})^\top$ and $a\in  [\underline{\alpha}, \overline{\alpha}]$,
\beql{num_var_multi_alpha_interpolation_network}
  \Sigma(a,y)=y_0+\sum_{j=1}^{o-1}  \boldsymbol{1}_{\alpha^{(j)}\leq z} \frac{( \alpha^{(j+1)}\wedge a -\alpha^{(j)})}{ \alpha^{(j+1)} -\alpha^{(j)}}y_j 
 .
\eeql
Algorithm \ref{alg:varsbis} thus produces
$$
\Sigma\big (\alpha,\zeta^{d, o}_{\lay +1}( X, \widehat{W}^{vars}) \big)\approx \widehat{q}
_{\alpha} (X) ,$$
where $ [\zeta^{d, o}_{\lay +1}( X, W)]_0$ can be interpreted as a predictor of the value-at-risk of lowest grid level $\alpha^{(1)}$, whereas, for each $j\ge 1$, $ [\zeta^{d, o}_{\lay +1}( X, W)]_j$ is a predictor of 
 the increment between the value-at-risks of levels $\alpha^{(j)}$ and $\alpha^{(j+1)}$.
Notice that one could impose the monotonicity by design by adding a positive activation function $\sigma$ to each neuron in the output layer of $\zeta^{d+1, o}_{\lay +1}$, except for the first neuron, e.g.~by replacing 
\bel & y_j \mbox{ with }\sigma(y_j)\mbox{, for all }j\in{1,\dots, o-1}, 
\eel
in \eqref{num_var_multi_alpha_interpolation_network}. However we have not found doing so to be satisfactory numerically and thus we keep the formulation in \eqref{num_var_multi_alpha_interpolation_network} as it is. In the numerics, we refer to this approach as {multi-$\alpha$(III)}.

\begin{algorithm}[!t] 
\small
\LinesNumbered
\SetAlgoLined
\SetKwInOut{AlgName}{name}
\SetKwInOut{Input}{input}\SetKwInOut{Output}{output}
\AlgName{MultiDiscreteVaRAlg\tcp{multi-$\alpha$(III)}}
\Input{$(X,Y)_{\nn}  $, a partition $\Batches$ of $\{1 \dots \then \} $, an increasing quantile level sequence $\alpha^{(1)}< \dots<\alpha^{(o)}$, a number of epochs $E\in\mathbb{N}^{\star}$, a learning rate $\eta>0$, initial values for the network parameters $\widehat{W} $, neural network output function $\zeta^{d, o}_{\lay +1}( x, W)$}
 
\Output{Trained parameters of multi-VaR network $\widehat{W}$} 
\tcp{Sample quantile levels $\alpha$}
$\alpha_i \sim \text{Uniform}(\underline{\alpha}, \overline{\alpha}) \quad \text{for } i = 1\dots \then$ \\
\tcp{Define a loss function}
define $\Sigma(y,a)=y_0+\sum_{j=1}^{o-1}  \boldsymbol{1}_{\alpha^{(j)}\leq a} \frac{( \alpha^{(j+1)}\wedge a -\alpha^{(j)})}{ \alpha^{(j+1)} -\alpha^{(j)}}y_j$

define $\displaystyle\pin^{vars} (W , \text{batch})=\frac{1}{|\text{batch}|}\sum_{i\in \text{batch}}  \Big[\Big(Y_i-\Sigma\big(  \zeta^{d, o}_{\lay +1}( X_i, W), \alpha_i\big)\Big)^++(1-\alpha_i)\Sigma\big(\zeta^{d, o}_{\lay +1}( X_i, W) ,\alpha_i  \big)\Big]$ 

$\widehat{W}^{vars} \leftarrow  \text{SGDOpt}(\{(X_i,Y_i)\}_{i = 1}^{\then}, \Batches, E, \eta, \widehat{W},\displaystyle\pin^{vars} )$
\caption{Learning multi discrete VaR.} 
\label{alg:varsbis}
\end{algorithm}

We now test the proposed procedures on a Student toy example and a dynamic initial margin (DIM) case study. Any minimization of loss functions over $\cNN(d,o,l,m,\sigma)$ or similar sets of neural networks is done using the Adam algorithm of \citet{kingma2014adam} over the parameters $W$ together with mini-batching.
 
All {of our} neural networks have $3$ hidden layers, and twice their input dimensionality as the number of neurons per hidden layer. In both examples below, for the multi-$\alpha$(I) and multi-$\alpha$(II) learning approaches, we use the bounds $(1-\overline{\alpha}, 1-\underline{\alpha})=({10}^{-4}, 0.15)$. For the multi-$\alpha$(III) approach, we use a uniform interpolation grid $1-\alpha^{(k)}={10}^{-3}+k\frac{0.15-{10}^{-3}}{20}$, with $k\in\{0, \dots, 20\}$. 
The different runs and the ensuing RMSE errors
referred to in our numerics share a common dataset. What is randomized from one run to the next is only the initialization of the network,
in order to
to make our numerical conclusions robust to the training noise related to the random initialization of the SGD.


\section{Conditionally Student-$t$ Toy Model}\label{sec:toymodel}
{\def\tilde{}

We first apply the above algorithms to the data generating process $(X,Y)$ such that $X$ is a standard multivariate normal vector 
\begin{align*}
    X \sim \cN(0, I_d)\mbox{,  for some } d\in\mathbb{N}^{\star} 
\end{align*}
and, for given functions $P, Q, S$ of $x$,
\begin{align}
    Y|X = T|X + UQ(X),
\end{align}
where $U$ is an independent Rademacher variable (worth $\pm 1$ with probabilities $1/2$), while the conditional distribution $T|X$ is Student-$t$ with degree $\nu>1$, location $P(X)$ and scale $S(X)$.
According to
\citet{khokhlov2016conditional},
\begin{align}
    \VaR(T|X) &= \Upsilon_{\nu}^{-1}(\alpha, P(X), S(X)) = P(X) + S(X) \Upsilon_{\nu}^{-1}(\alpha), \\
    \ES(T|X) &= P(X) + S(X) \frac{\den _\nu(\Upsilon_\nu^{-1}(\alpha))}{1-\alpha}\frac{\nu + (\Upsilon_\nu^{-1}(\alpha))^2}{\nu - 1},
\end{align}
where $\Upsilon_\nu(\cdot, \mu, \sigma)$ and $\den _\nu(\cdot, \mu, \sigma)$, shortened  for $\mu = 0$ and $\sigma = 1$ as  $\Upsilon_\nu(\cdot )$ and $\den _\nu(\cdot )$, are the cdf and pdf of the Student-$t$ distribution with degree $\nu$, location $\mu$ and scale $\sigma$. 
The cdf of $Y|X$ is given by
\beql{eq:t_cdf}
    \tilde{\Upsilon}_{Y|X} (y) &= \PP(Y<y|X) = \frac{1}{2}\Big[\PP(Y<y|X, U = 1)+  \PP(Y<y|X, U = -1)\Big] \\
    &= \frac{1}{2}\Big[\PP(T<y+Q(X)) + \PP(T<y-Q(X)) \Big] \\
    &= \frac{1}{2}\Big[\Upsilon_\nu(y, P(X) - Q(X), S(X)) + \Upsilon_\nu(y, P(X) + Q(X), S(X)) \Big],
    \eeql
with corresponding conditional pdf 
\begin{align}\label{eq:t_pdf}
    \tilde{\den }_{Y|X}(y) = \frac{1}{2}\Big[\den _\nu(y, P(X)-Q(X), S(X)) + \den _\nu(y, P(X)+Q(X), S(X))\Big].
\end{align}
The conditional VaR of $Y$ given $X$, at level $\alpha$, by
\begin{align*}
\VaR(Y|X) = \tilde{\Upsilon}^{-1}_{Y|X} (\alpha)    ,
\end{align*}
cannot be  computed analytically, but one can efficiently approximate it  numerically (by monotonicity of quantile functions). 
The formula for the ES of $Y|X$ is deduced by the following result, whose proof is deferred to Section \ref{e:sk}:
\begin{align}
    &\ES(Y|X)= P(X) +  \frac{1}{2(1-\alpha)}Q(X)\left[\Upsilon_\nu\left(\tilde{\Upsilon}^{-1}_{Y|X} (\alpha), P(X)-Q(X), S(X)\right)\right. \\
    &\qqq \qqq\left.-\Upsilon_\nu\left(\tilde{\Upsilon}^{-1}_{Y|X} (\alpha), P(X)+Q(X), S(X)\right)\right] \\
    &\qqq+ \frac{\nu S(X)^2+ (\tilde{\Upsilon}^{-1}_{Y|X} (\alpha)-P(X))^2+ Q(X)^2}{(\nu -1)(1-\alpha)}\tilde{\den }_{Y|X}\left(\tilde{\Upsilon}^{-1}_{Y|X} (\alpha) \right) \\
    &\qqq+ \frac{ \left(\tilde{\Upsilon}^{-1}_{Y|X} (\alpha) -P(X)\right)Q(X)}{(\nu -1)(1-\alpha)}\left[\den _\nu\left(\tilde{\Upsilon}^{-1}_{Y|X} (\alpha), P(X)-Q(X), S(X) \right) \right.\\
    &\qqq \qqq\left.- \den _\nu\left(\tilde{\Upsilon}^{-1}_{Y|X} (\alpha), P(X)+Q(X), S(X) \right)\right]. \label{e:ESt}
\end{align}

\noindent
The corresponding values of
$\VaR(Y|X)$ and
$\ES
(Y|X)$
 will serve us as ground-truth values. 
This provides an heavy-tailed setup in which the solution can be computed quasi analytically, for benchmarking purposes, but sufficiently rich so that the data are not parametrically determined by a few scaling parameters (as it would be the case with conditionally Gaussian or even elliptical distributions).

}

\subsection{Numerical Results}

We take $P, Q$, and $S$ as quadratic functions of $x$ and set $\nu = 3$.
We use a dimension of $d=25$ for the state space of $X$.
The nonzero coefficients %
of these polynomials are drawn independently from a standard normal distribution. For this example, we use $n=2^{19}=524288$ training points and the same number of testing points for computing the errors. For the Adam algorithm, we used $2000$ epochs, mini-batching with batches of size $2^{15}=32768,$ a learning rate $\eta=0.01$, and the rest of the parameters kept at their default values as per
 \citep{kingma2014adam}.
 
Figures \ref{fig:converge_NRMSE_var_es} shows the convergence of the learnings at rate close to $n^{-\frac{1}{4}}$, consistent with the controls  of Theorem  \ref{t:cvrateNNq} for the VaR and of Theorem  \ref{t:cvesnn} for the ES. As our sample size $n =2^{19}$ satisfies \eqref{equdesineequ:thresold}, this is in line with the discussion following Theorem \ref{thebouerrrrad}. Notice that setting $\nu = 3$ yields the integrability of order $p=2$ of the  $t$-distributed response $Y$, making the statistical bound in \citet{shen2021deep} have a lower rate than ours (see Remark \ref{r:rate}).

\begin{figure}[!tp]
\begin{subfigure}{0.5\linewidth}
  \centering
  \includegraphics[width=\linewidth]{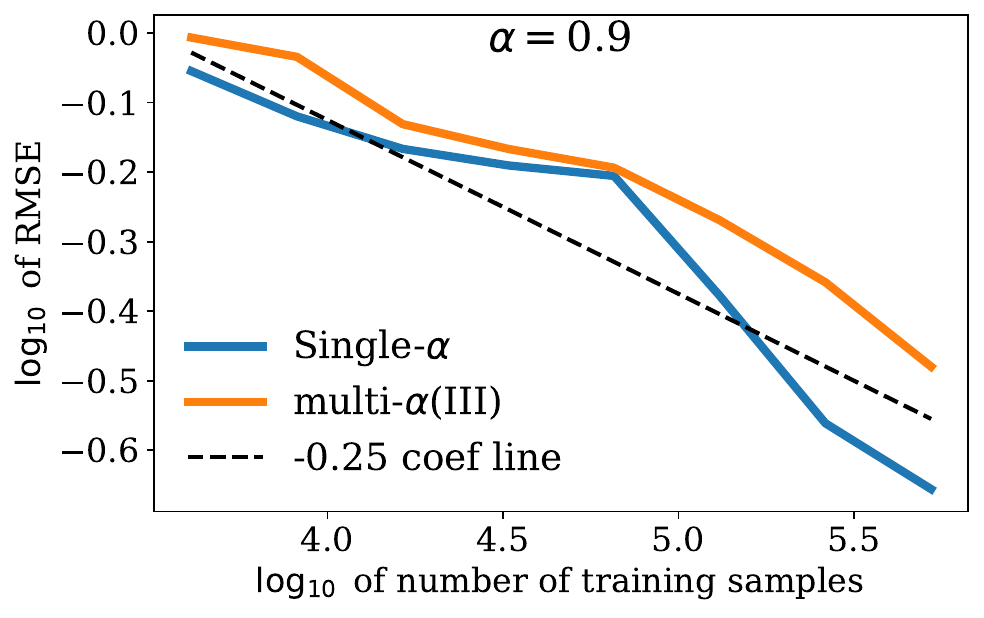}
\end{subfigure}%
\hspace*{-0.2cm}
\begin{subfigure}{.5\linewidth}
  \centering
  \includegraphics[width=\linewidth]{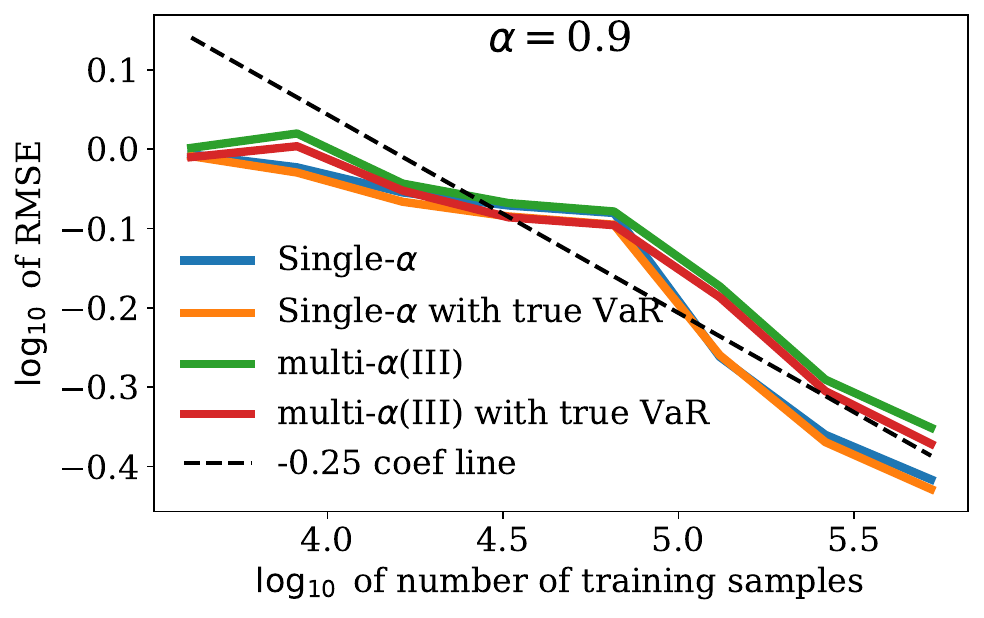}
  
\end{subfigure}

\begin{subfigure}{.5\linewidth}
  \centering
  \includegraphics[width=\linewidth]{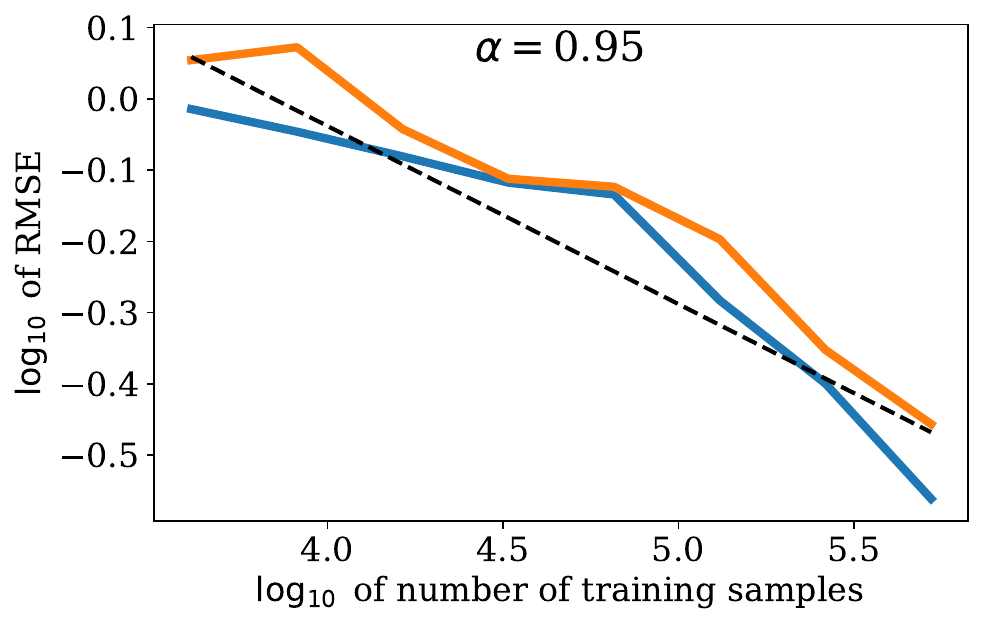}
\end{subfigure}%
\hspace*{-0.2cm}
\begin{subfigure}{.5\linewidth}
  \centering \includegraphics[width=\linewidth]{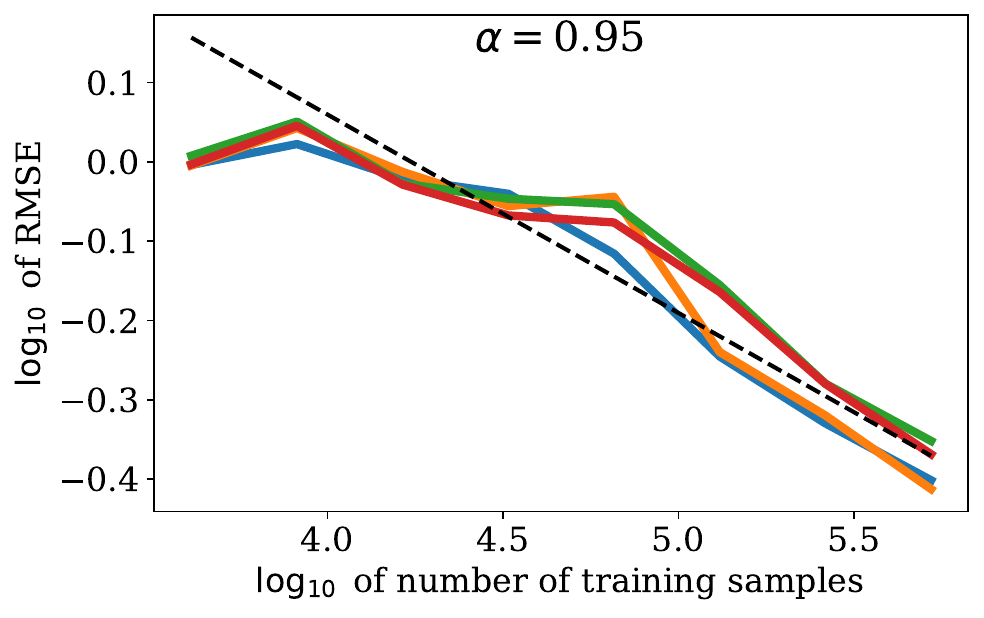}
\end{subfigure}

\begin{subfigure}{.5\linewidth}
  \centering
  \includegraphics[width=\linewidth]{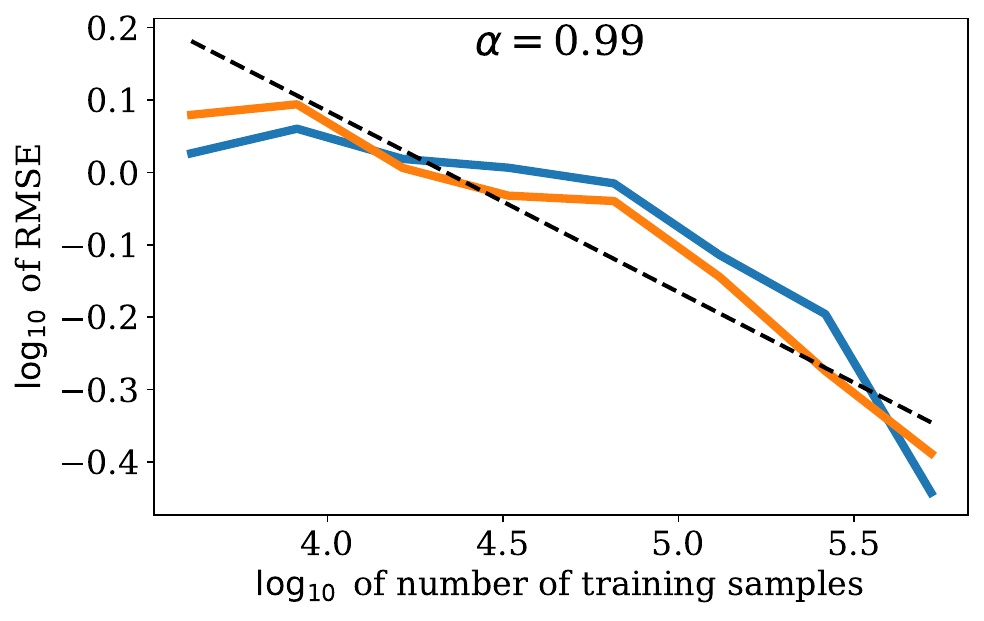}
\end{subfigure}%
\hspace*{-0.2cm}
\begin{subfigure}{.5\linewidth}
  \centering \includegraphics[width=\linewidth]{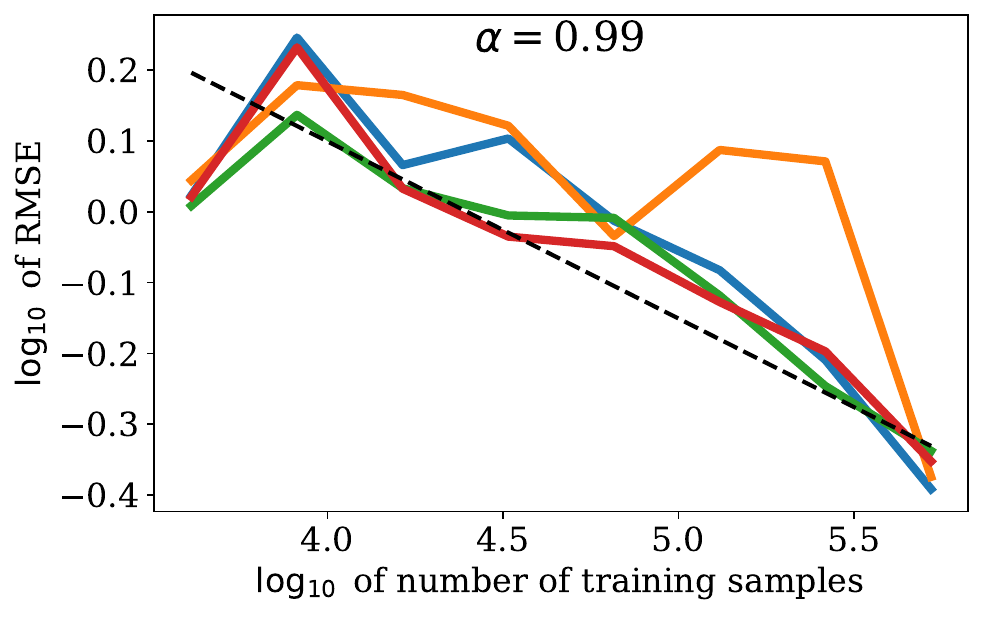}
\end{subfigure}

\begin{subfigure}{.5\linewidth}
  \centering
  \includegraphics[width=\linewidth]{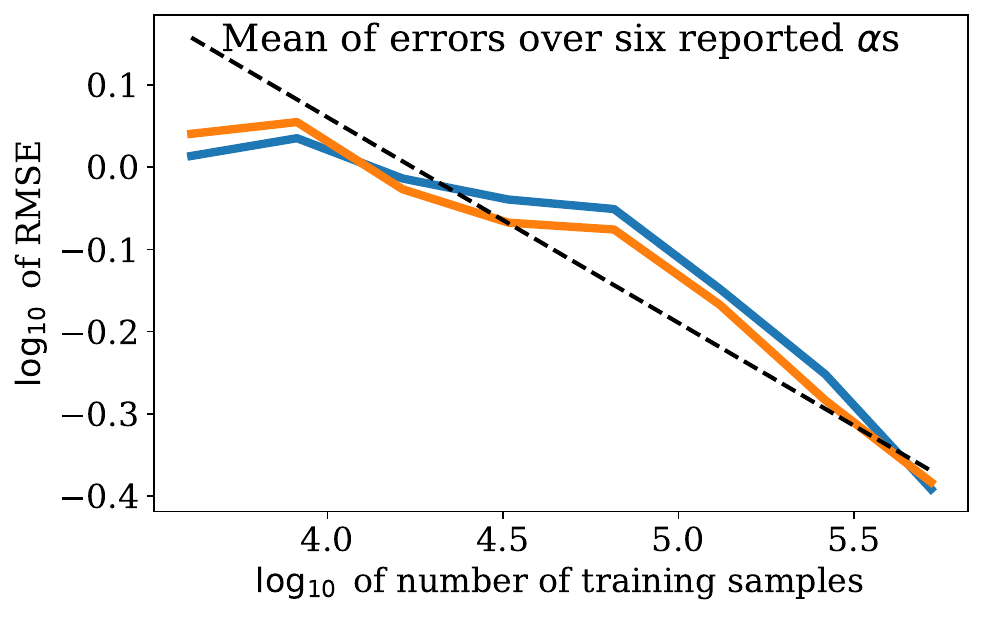}
\end{subfigure}%
\hspace*{-0.2cm}
\begin{subfigure}{.5\linewidth}
  \centering \includegraphics[width=\linewidth]{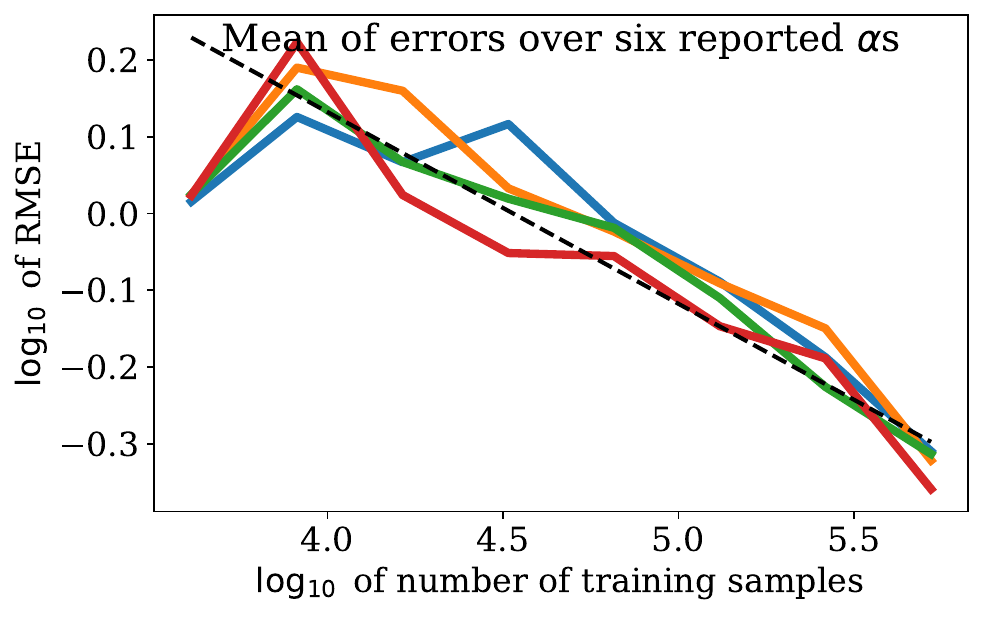}
\end{subfigure}

\caption{Convergence curves of the log-RMSE errors against their groundtruth values, in the Student-$t$ toy-example when increasing (the log of) the number of training samples, of:(\textit{Left}) the conditional VaR learned by the single and multi-$\alpha$(III) models;  (\textit{Right}) the conditional ES learned by the single and multi-$\alpha$(III) models, using the true or learned conditional VaR as VaR hypothesis for the ES. The slope of the black dash lines correspond to the theoretical convergence rate of 
 Theorems \ref{t:cvrateNNq} and \ref{t:cvesnn}.} 
\label{fig:converge_NRMSE_var_es}
\end{figure}
 
Tables \ref{tbl:student_var_rmse}, as also \ref{tbl:IM_25_var_rmse}, \ref{tbl:IM_50_var_rmse} and \ref{tbl:IM_75_var_rmse} below in the DIM case, suggest that the multi-$\alpha$ approaches are competitive compared to the single-$\alpha$ approach by yielding acceptable errors for confidence levels below $99\%$, while requiring only one single training, as opposed to the single-$\alpha$ approach which requires one training per target confidence level. 
For very extreme confidence levels, like $99.9\%$, the {multi-$\alpha$(III)} approach outperforms all the other approaches. This can be explained by the fact that, even if the target confidence level is hard to reach given a limited training set, the lower
confidence levels in the interpolation grid contribute to inferring the VaR for the target confidence level. Table~\ref{tbl:student_pvalue} confirms that one can rely on the twin-simulation trick of Section \ref{ss:trick} to draw mostly similar conclusions as in Table \ref{tbl:student_var_rmse}, without the need to have access to the groundtruth estimators. Note that we computed upper-bounds of 95\% confidence intervals for \eqref{eq:twinsim_var_err},
instead of the estimates directly in order to be conservative and take into account the potentially high variance in the indicator functions that need to be simulated in order to estimate \eqref{eq:twinsim_var_err}. Table~\ref{tbl:student_crossing} demonstrates the effectiveness of the penalization term (for $\lambda$ simply set to 1) in the multi-$\alpha$(I) approach to mitigate the quantiles crossing problem. Table~\ref{tbl:student_crossing} also shows that the other multi-$\alpha$ learning approaches, even without directly penalizing the crossing of the quantiles, behave better than a single-$\alpha$ learning in terms of the crossing of the quantiles.
\FloatBarrier
\begin{table}[!t]
\centering
\begin{tabular}{lrrr}
\toprule
$\alpha$       & 0.999                  & 0.995                  & 0.99                   \\ \midrule
single-$\alpha$     & 0.882 (0.037)          & 0.548 (0.035)          & 0.429 (0.016)          \\
multi-$\alpha$(I)   & 0.821 (0.048)          & 0.438 (0.043)          & \textbf{0.342 (0.035)} \\
multi-$\alpha$(II)  & 0.81 (0.05)            & 0.438 (0.035)          & 0.352 (0.034)          \\
multi-$\alpha$(III) & \textbf{0.549 (0.037)} & \textbf{0.422 (0.028)} & 0.408 (0.022)          \\
 \bottomrule
\end{tabular}
\begin{tabular}{lrrr}
\toprule
$\alpha$            & 0.98                   & 0.95                  & 0.9                    \\ \midrule
single-$\alpha$     & 0.363 (0.014)          & \textbf{0.25 (0.013)} & \textbf{0.209 (0.013)} \\
multi-$\alpha$(I)   & \textbf{0.319 (0.035)} & 0.263 (0.02)          & 0.242 (0.014)          \\
multi-$\alpha$(II)  & 0.325 (0.036)          & 0.26 (0.019)          & 0.237 (0.014)          \\
multi-$\alpha$(III) & 0.394 (0.029)          & 0.346 (0.031)         & 0.34 (0.034)           \\
 \bottomrule
\end{tabular}
\caption{Means (standard deviations) of RMSE errors of learned conditional VaR estimators against groundtruth values in the Student-$t$ toy-example across 32 runs. The RMSE errors are normalized by division by the standard deviation of the groundtruth VaR.}
\label{tbl:student_var_rmse}
\end{table}

\begin{table}[!t]
\centering
\begin{tabular}{lrrr}
\toprule
$\alpha$      & 0.999           & 0.995           & 0.99            \\ \midrule
single-$\alpha$    & -1.989 (-2.842) & -1.641 (-2.756) & -1.528 (-2.716)          \\
multi-$\alpha$(I)  & -2.029 (-2.828) & -2.0 (-2.846)   & \textbf{-1.903 (-2.906)} \\
multi-$\alpha$(II) & -2.019 (-2.846) & -1.98 (-2.893)  & -1.888 (-2.834)          \\
multi-$\alpha$(III) & \textbf{-2.36 (-2.684)}   & \textbf{-2.11 (-2.699)}   & -1.713 (-2.478)          \\
 \bottomrule
\end{tabular}
\begin{tabular}{lrrr}
\toprule
$\alpha$   & 0.98                   & 0.95                  & 0.9     \\ \midrule
single-$\alpha$     & -1.399 (-2.83)  & -1.311 (-2.549) & \textbf{-1.199 (-2.68)} \\
multi-$\alpha$(I) & \textbf{-1.691 (-2.908)}  & \textbf{-1.351 (-2.622)}  & -1.159 (-2.538)          \\
multi-$\alpha$(II)  & -1.676 (-2.762) & -1.349 (-2.662) & -1.159 (-2.525)         \\
multi-$\alpha$(III) & -1.584 (-2.412) & -1.273 (-2.309) & -1.051 (-2.222)         \\
 \bottomrule
\end{tabular}
\caption{Means (standard deviations) log$_{10}$ of the means (standard deviations)  across 32 runs of a posteriori twin Monte Carlo {$p$-value  error estimates}, i.e.~of the right-hand side of \eqref{eq:twinsim_var_err}, of learned conditional VaR estimators in the Student-$t$ toy-example. 
}
\label{tbl:student_pvalue}
\end{table}

\begin{table}[htbp]
\centering
\resizebox{\textwidth}{!}{
\begin{tabular}{lrrr}
\toprule
{$E$} &  $\{\empappvar_{0.999}(X)<\empappvar_{0.995}(X)\}$ &  $\{\empappvar_{0.995}(X)<\empappvar_{0.99}(X)\}$ &  $\{\empappvar_{0.99}(X)<\empappvar_{0.98}(X)\}$
\\
\midrule
single-$\alpha$     & -0.817 (-1.871)          & -0.601 (-1.85)           & -0.706 (-1.861)          \\
multi-$\alpha$(I)   & \textbf{-4.692 (-4.311)} & \textbf{-4.412 (-4.156)} & \textbf{-3.943 (-3.869)} \\
multi-$\alpha$(II)  & -4.449 (-4.094)          & -4.209 (-3.97)           & -3.79 (-3.72)            \\
multi-$\alpha$(III) & -3.445 (-2.986)          & -3.346 (-2.959)          & -1.674 (-1.852)          \\
\bottomrule
\end{tabular}
}
\resizebox{\textwidth}{!}{
\begin{tabular}{lrrr}
\toprule
{$E$} &   $\{\empappvar_{0.98}(X)<\empappvar_{0.97}(X)\}$ &  $\{\empappvar_{0.97}(X)<\empappvar_{0.96}(X)\}$ &  $\{\empappvar_{0.96}(X)<\empappvar_{0.95}(X)\}$ \\
\midrule
single-$\alpha$     & -0.589 (-1.687)          & -0.519 (-1.511)          & -0.488 (-1.601)          \\
multi-$\alpha$(I)   & \textbf{-3.219 (-3.315)} & \textbf{-2.516 (-2.626)} & -1.929 (-2.076)          \\
multi-$\alpha$(II)  & -3.132 (-3.189)          & -2.488 (-2.648)          & \textbf{-1.936 (-2.139)} \\
multi-$\alpha$(III) & -1.824 (-1.994)          & -1.39 (-1.585)           & -1.721 (-1.766)          \\
\bottomrule
\end{tabular}}
\caption{log$_{10}$ of the empirical estimates (and of the corresponding standard deviations) of $\PP(E)$, for the events $E$ listed in the first row, for learned conditional VaR estimators in the Student-$t$ toy-example across 32 runs.  }
\label{tbl:student_crossing}
\end{table}

For the ES learning in the  Student-$t$ toy-example, for brevity, we denote by {``LR using M VaR"} an ES learning using linear regression only for the output layer, corresponding to the linear regression case in Algorithm \ref{alg:es}, and a VaR learned using the method $M$ as the candidate VaR. For example, LR using single-$\alpha$ VaR refers to the linear regression approach for learning the ES, by using a VaR that is learned with the {single-$\alpha$} approach as the VaR candidate. To demonstrate the effectiveness of this linear regression approach, we also introduce an ES that is learned by neural regression, by using a neural network corresponding to the second (else) case in Algorithm \ref{alg:es}, without freezing any weights  
 and using the groundtruth VaR as the VaR candidate. Table~\ref{tbl:student_ES_rmse} shows that our linear regression approach for the ES outperforms the neural regression, no matter which approach is used for learning the embedded VaR candidate. The relative performance of the different linear regression approaches in Table~\ref{tbl:student_ES_rmse} is explained by the relative performance of the VaR learning approaches, given that the VaR learning error contributes to the ES learning error.

\begin{table}[!t]
\centering
\begin{tabular}{lrrr}
\toprule
{$\alpha$} &  0.999 &  0.995 &  0.99\\
\midrule
NNR using true VaR              & 1.598 (0.196) & 1.063 (0.045) & 1.006 (0.022) \\
LR using single-$\alpha$ VaR     & \textbf{0.955 (0.015)}   & 2.231 (1.149)            & 0.804 (0.429)           \\
LR using multi-$\alpha$(I) VaR  & 2.581 (0.505) & 0.799 (0.169) & 0.55 (0.098)  \\
LR using multi-$\alpha$(II) VaR & 2.394 (0.396) & 0.74 (0.137)  & 0.518 (0.081) \\
LR using multi-$\alpha$(III) VaR & 1.276 (0.355)            & \textbf{0.574 (0.085)}   & \textbf{0.491 (0.042)}  \\
\bottomrule
\end{tabular}
\begin{tabular}{lrrr}
\toprule
{$\alpha$} &  0.98 &  0.95 &  0.9 \\
\midrule
NNR using true VaR               & 0.968 (0.014) & 0.887 (0.029) & 0.816 (0.079) \\
LR using single-$\alpha$ VaR    & 0.492 (0.091)            & \textbf{0.371 (0.021)}   & \textbf{0.383 (0.027)}  \\
LR using multi-$\alpha$(I) VaR   & 0.43 (0.053)  & 0.38 (0.027)  & 0.403 (0.02)  \\
LR using multi-$\alpha$(II) VaR & \textbf{0.416 (0.046)}   & 0.374 (0.024)            & 0.396 (0.017)           \\
LR using multi-$\alpha$(III) VaR & 0.45 (0.029)  & 0.432 (0.025) & 0.432 (0.02)  \\
\bottomrule
\end{tabular}
\caption{Means (standard deviations) of RMSE errors of learned conditional ES estimators against groundtruth values in the Student-$t$ toy-example across 32 runs. The RMSE errors are normalized by division by the standard deviation of the groundtruth ES.}
\label{tbl:student_ES_rmse}
\end{table}

\FloatBarrier
\section{Dynamic Initial Margin Case Study\label{sec:imstoprocess}}
\emmi{this section has only one subsection: a bit weird. Either remove the splitting or add another subection.}
A financial application of the quantile learning framework is the learning of a path-wise, dynamic initial margin (DIM) in the context of XVA computations (see e.g.\ \citeN[Table 4.1]{Crepey21} and \citet[Section A.4]{albanese2021xva}). Let there be given respectively $\mathbb{R}^d$  valued  and real valued stochastic processes $S=(S_t)_{t\geq 0}$ and $\text{MtM}=(\text{MtM}_t)_{t\geq 0}$, where $S$ is Markov and  $S_t$ represents the state of the market at time $t$ (e.g.~diffused market risk factors),  whereas $\text{MtM}_t$ represents the mark-to-market (price) of the portfolio of the bank at time $t$. We include in this price the cash flows cumulated up to time $t$, so that $\text{MtM}_{t+\delta}-\text{MtM}_t$ is $\sigma(X_s, t \leq s \leq t+\delta)$ measurable. We ignore risk-free discounting in the notation (while preserving it in the numerical experiments). The initial margin of the bank at time $t$ at the confidence level $\alpha$, denoted by $\text{IM}_t$, is defined as 
\begin{equation}
\label{eqdefim}
  \text{VaR}
\left(\text{MtM}_{t+\delta}-\text{MtM}_t\left|S_t\right.\right)
\defeq \text{IM}
 (t,S_t).
\end{equation}
Hence, having $n$ Euler simulated paths of $S$ and $\text{MtM}$, one can estimate the DIM process $(\text{IM}(t,S_t))$ at grid times, using one quantile regression with data
$(X,Y)_{\nn}\equiv (S_t,\text{MtM}_{t+\delta}-\text{MtM}_t)_{\nn}$
for each grid time $t$.
Alternatively, this DIM process can be estimated by a brute force nested Monte Carlo method detailed in the arXiv v1 preprint version of this work, taking several nights of computation time on a workstation, as opposed to a few minutes by regressions.

\FloatBarrier

\subsection{Numerical Results}
  
We consider a portfolio composed of $100$ interest rate swaps with randomly drawn characteristics and final maturity 10 years, assessed in the market model of \citet*[Appendix B]{hierarchicalsim2021}, i.e.\
 a multi-factor market model with $10$ short-rate processes representing $10$ economies and $9$ cross-currency rate processes. Given  %
that swap coupons can depend on short-rates at previous fixing dates, we also include in the regression basis the same short-rates but observed at the latest previous fixing date, which leads in total to a dimensionality of $d=29$ for the state vector $S_t$ at a given time $t>0$, with $100$ time steps uniformly spread between time 0 and the final maturity of the portfolio equal to 10 years. We use $2^{22}=4194304$ simulated paths (generated in 25 seconds using the code developed in \citeN[Appendix B]{hierarchicalsim2021}) for training and $2^{14}$ simulated paths, independent of the former, for evaluating the nested Monte Carlo benchmark and computing the errors. We leverage the transfer learning trick used in \citet[Appendix B]{hierarchicalsim2021}, which consists in doing the training starting from the latest time-step and then proceeding backwards by reusing the solution obtained at each successive time-step $t_{k+1}$ as an initialization for the learning to be done at time $t_k$. This allows us to use only $16$ training epochs. As in the Student toy-example, we use mini-batching. The batch size is taken to be $2^{17}=131072$, we use a learning rate of $0.001$, and the rest of the Adam parameters are kept at their default values.

To illustrate that the quantile learning approach allows one to learn an entire stochastic process (dynamic initial margin), we plot the mean and $5$  th/$95$  th percentiles of the learned IM process at each time-step for the different quantile learning schemes in Figure~\ref{fig:IM_all}. 
At $t=0$ the IM is deterministic because there is no randomness in the model yet, at $t=10$ (last time step, i.e.\ final maturity of the portfolio) 
the IM vanishes because there are no later cash flows. 
The {sawtooth}-like behaviour in the paths of $(\text{IM}
_t)_{t\geq 0}$ that is visible in the plots in Figure~\ref{fig:IM_all} 
is expected, due to the recurring cash-flows inherent to interest rate swaps
\citep*{AndersenPykhtinSokol17}. %
\begin{figure}[!bp]
\centering
\includegraphics[width=1.02\linewidth]{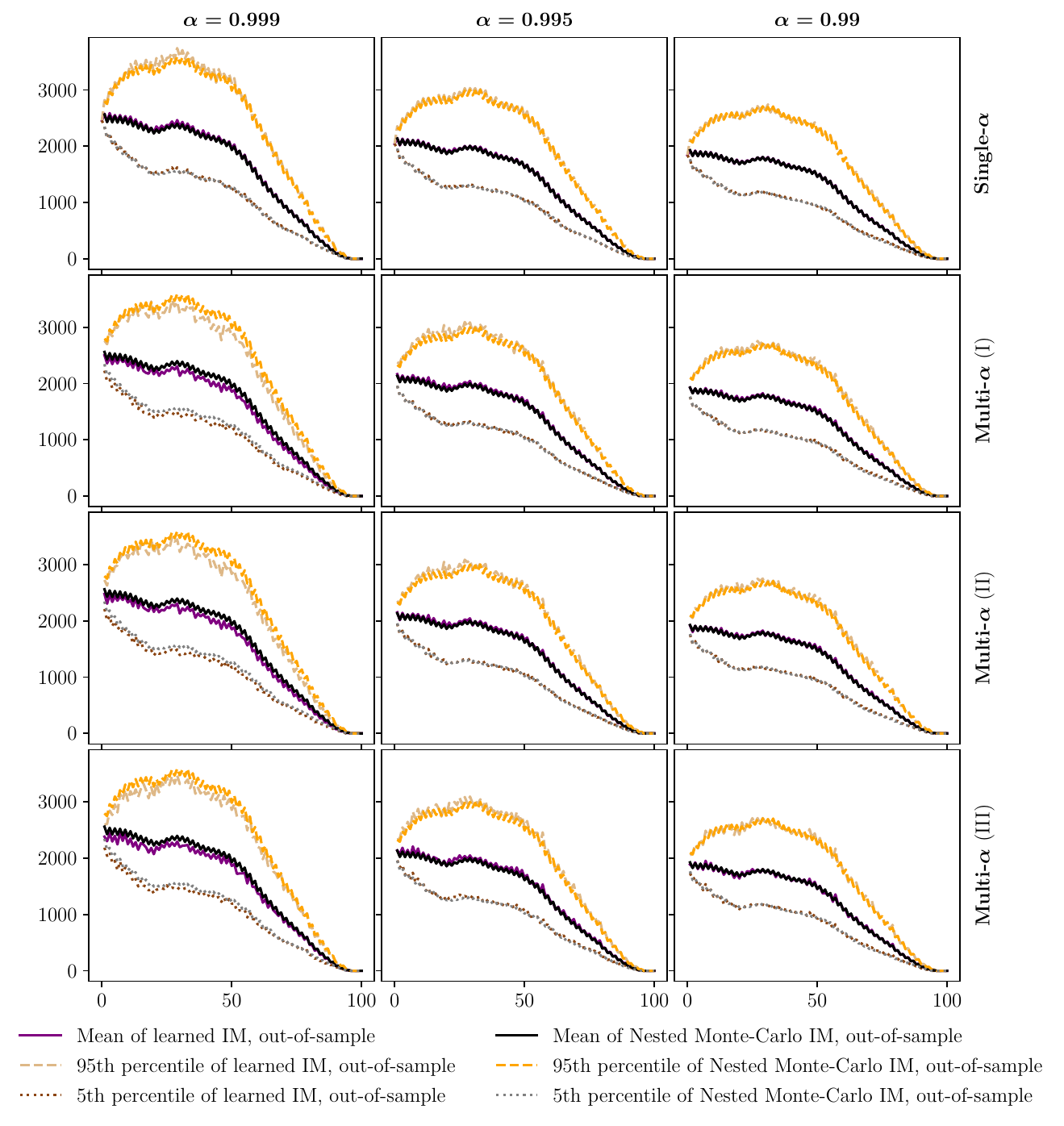}
\caption{Mean and $5$  th/$95$  th percentiles of both the learned and the nested Monte Carlo IM at different time steps and for different values of $\alpha$ and learning approaches. The learning approach used for the plots in each row is indicated on the right, and each column corresponds to one value of $\alpha$ which is indicated at the top of each column. Statistics are computed using out-of-sample 
trajectories of the diffused risk-factors, and the time steps are on the $x$  axis.}
\label{fig:IM_all}
\end{figure}
\FloatBarrier
Tables \ref{tbl:IM_25_var_rmse}, \ref{tbl:IM_50_var_rmse} and \ref{tbl:IM_75_var_rmse} (using the nested Monte Carlo as a benchmark) confirm the conclusions of Table \ref{tbl:student_var_rmse} regarding the competitiveness of the multi-$\alpha$ approaches.
\begin{table}[!tp]
\centering
\begin{tabular}{lrrrrrr}
\toprule
{$\alpha$} &  0.999 &  0.995 &  0.99 &  0.98 &  0.95 &  0.9 \\
\midrule
multi-$\alpha$(I)   &  0.265 &  0.160 &  0.109 &  0.065 &  0.058 &  {\bfseries 0.056} \\
multi-$\alpha$(II)  &  0.261 &  0.155 &  0.107 &  0.066 &  {\bfseries 0.057} &  {\bfseries 0.056} \\
multi-$\alpha$(III) &  {\bfseries 0.128} &  0.185 &  0.102 &  0.133 &  0.116 &  0.074 \\
Single-$\alpha$      &  0.134 &  {\bfseries 0.074} &  {\bfseries 0.070} &  {\bfseries 0.056} &  0.066 &  0.065 \\
\bottomrule
\end{tabular}
\caption{RMSE errors of learned $\text{IM}_t$ estimators against nested Monte Carlo estimators, for $t=2.5 \text{years}$. Errors are normalized by division by the standard deviation of the nested Monte Carlo benchmark.}
\label{tbl:IM_25_var_rmse}
\end{table}\begin{table}[!tp]
\centering
\begin{tabular}{lrrrrrr}
\toprule
{$\alpha$} &  0.999 &  0.995 &  0.99 &  0.98 &  0.95 &  0.9 \\
\midrule
multi-$\alpha$(I)   &  0.204 &  0.166 &  0.131 &  0.072 &  0.061 &  0.069 \\
multi-$\alpha$(II)  &  0.212 &  0.162 &  0.127 &  0.072 &  0.062 &  0.069 \\
multi-$\alpha$(III) &  {\bfseries 0.150} &  0.123 &  {\bfseries 0.067} &  0.065 &  0.066 &  0.068 \\
Single-$\alpha$      &  0.165 &  {\bfseries 0.095} &  0.070 &  {\bfseries 0.057} &  {\bfseries 0.060} &  {\bfseries 0.066} \\
\bottomrule
\end{tabular}
\caption{RMSE errors of learned $\text{IM}_t$ estimators against nested Monte Carlo estimators, for $t=5 \text{years}$. Errors are normalized by division by the standard deviation of the nested Monte Carlo benchmark.}
\label{tbl:IM_50_var_rmse}
\end{table}\begin{table}[!tp]
\centering
\begin{tabular}{lrrrrrr}
\toprule
{$\alpha$} &  0.999 &  0.995 &  0.99 &  0.98 &  0.95 &  0.9 \\
\midrule
multi-$\alpha$(I)   &  0.292 &  0.119 &  0.122 &  0.095 &  0.073 &  0.072 \\
multi-$\alpha$(II)  &  0.296 &  0.118 &  0.118 &  0.091 &  0.071 &  0.070 \\
multi-$\alpha$(III) &  0.157 &  0.118 &  0.090 &  0.089 &  0.079 &  0.086 \\
Single-$\alpha$      &  {\bfseries 0.119} &  {\bfseries 0.088} &  {\bfseries 0.082} &  {\bfseries 0.068} &  {\bfseries 0.061} &  {\bfseries 0.061} \\
\bottomrule
\end{tabular}
\caption{RMSE errors of learned $\text{IM}_t$ estimators against nested Monte Carlo estimators, for $t=7.5 \text{years}$. Errors are normalized by division by the standard deviation of the nested Monte Carlo benchmark.}
\label{tbl:IM_75_var_rmse}
\end{table}

\section*{Conclusion of the Numerical Experiments}
The numerical experiments of Sections \ref{sec:toymodel} and \ref{sec:imstoprocess} suggest that learning multiple quantiles (multi-$\alpha$(I), multi-$\alpha$(II) or multi-$\alpha$(III)), although counter-intuitive at first, can help better target extreme quantiles than a standard single quantile learning approach.  This can be explained by the fact that multiple quantile approaches leverage the information given by nearby quantiles and thus are better at extrapolating at the extremes. The multi-$\alpha$(I) approach is remarkably good at ensuring, via soft-constraints on the derivative with respect to the quantile level, monotonicity (avoiding quantile crossings), in cases where consistency among different quantile levels is desired. Our experiments also show that one can successfully use these quantile estimation methods in an XVA or dynamic risk calculation setting, where the computation burden may be greatly diminished by replacing nested Monte Carlo estimations by VaR and ES learnings.

As \textbf{practical take-away messages} to the reader, we would like to emphasize three algorithmic breakthroughs of the paper, namely
\textbf{(i)} the a posteriori twin Monte Carlo validation trick of Algorithm \ref{alg:twin}, \textbf{(ii)}  the neural network VaR to ES transfer learning trick corresponding to the linear regression case in Algorithm  \ref{alg:es}, and 
\textbf{(iii)} the multi-$\alpha$ learning schemes   Algorithms \ref{alg:vars}-\ref{alg:varsbis}.

\appendix 
\section{Value-at-Risk and Expected Shortfall Representations\label{s:qsrepr}} 
In this section we recall the  elicitability results behind our
learning algorithms.

A cumulative distribution function (cdf) $F:\R\to [0,1]$ is by definition (Stieltjes) integrable if 
\begin{align}
\label{equdefdisint}
\int_{\R}|y| \,F(dy)<\infty .
\end{align}
If $Y$ is a random variable 
with \eg{cdf} $F$ (i.e.~$\PP[Y\leq t]=F(t),$ $t\in \R$), then \eqref{equdefdisint} holds if and only if $Y$ is $\PP$  integrable (the left-hand side of \eqref{equdefdisint} is then $\PE|Y|$). 

\bd
Let $F:\R\to [0,1]$ be an integrable cdf and let $\alpha\in (0,1)$. The {\upshape value-at-risk} (\VaR) and {\upshape expected shortfall} (\ES) of $F$ at the confidence level $\alpha$ are defined respectively by
\begin{align}
\VaR(F)\eqd \min F^{-1}([\alpha,1])\sp \ES_{}(F)=\frac{1}{1-F(\VaR(F){-})}\int_{[\VaR(F),\infty)}y \,F(dy). 
\label{equdefvares}
\end{align}
If $Y$ is an integrable random variable on the probability space $(\Omega,\cA,\PP)$, we write 
\begin{align}
\VaR(Y)\eqd \VaR(F_{Y})\sp  \ES(Y)\eqd \ES(F_{Y}),
\end{align}
where $F_{Y}(t)=\PPro{Y\leq t}$ is the \eg{cdf} of $Y$.\fe 
\eds
\brem
If $Y$ is an integrable random variable,
then it is easy to see that
\begin{align}
\label{equexiqua}
\VaR
(Y)= \min \{t:\PPro{Y\leq t}\geq \alpha\}, &
\qquad \ES
(Y)=\Esp{Y|Y\geq \VaRa (Y)}
\end{align}  
The version of \eqref{equexiqua}
 for abstract distribution functions $F$ is clear {\it mutatis mutandis}.\fe 
\erems
 

\bhyp
\label{assconfarvar}
There exists an interval $[a,b]$ where $F$ is continuous and
\begin{align*}
F(a)< \alpha < F(b).\fe
\end{align*}
\ehyp

Modulo the additional truncation feature needed for this work,
The following result corresponds to \citet{rock:urya:00}.
\bl
\label{theelivartil}
Let $\lpvar_{\hun}$ be given by \eqref{equdefinslosvar}.
If $F$ is an integrable distribution function satisfying Assumption \ref{assconfarvar}, then 
\begin{align}
\begin{split}
F^{-1}(\{\alpha\}) &= \argmin_{u\in [a,b]} \underbrace{(1-\alpha)^{-1}\int_{\theu}^{\infty}( y - \theu )^{+}F(dy)+\theu}_{\defeq es(u)} \\
&= \argmin_{u\in [a,b]}\underbrace{(1-\alpha)^{-1}\int_{\theu}^{\infty}( (a \vee y\wedge b) - \theu )^{+}F(dy)+\theu}_{\defeq es^{a,b}(u)} 
\end{split}
\label{equchavarmin}   \\
\label{equchaexpshomin}
\{\ES
(F)\}&= es \,( F^{-1}(\{\alpha\})\,) 
. 
\end{align}

\el

\proof Under Assumption \ref{assconfarvar}, $\O   \subsetneq  F^{-1}(\{\alpha\})\subseteq (a,b) $.
Since $F$ is  continuous in $[a,b]$, the identity 
$\partial_\theu es (u)
= 1-(1-\alpha)^{-1}(1-F(\theu))
$
holds for $\theu\in [a,b]$. It follows in particular that the continuously differentiable function $es$ has critical points as the set of $\alpha$-quantiles of $F$. Since $es$ is convex,
these critical points are the minimizers of $es$ over $[a,b]$, which proves the first equality in \eqref{equchavarmin}. Likewise,   
\beql{e:ltab}
{\partial_\theu}es^{a,b} (u)&= 1 - (1-\alpha)^{-1}\Big(((a \vee u\wedge b) - u)^+ + \int_{u}^\infty \textbf{1}_{(a \vee y\wedge b) \geq u} F(dy)\Big)\\
&=
    1 -  (1-\alpha)^{-1}(1- F(u)) \sp \mbox{for $u\in [a,b]$}.
\eeql
Consequently, $F^{-1}(\{\alpha\})$ are the critical points of the function under the second argmin in \eqref{equchavarmin}, which are also the minimizers over $[a,b]$ due to the convexity of $es^{a,b}$. This completes the proof of  \eqref{equchavarmin}.

Moreover, for any $q\in F^{-1}(\{\alpha\})$, $F$ is constant on $[\VaR(F),\qalpha)$, hence
\eqref{equdefvares} yields
\begin{align}
\ES
(F)={(1-\alpha)}^{-1}\int_{\qalpha}^{\infty}y F(dy)= es(u)
. 
\end{align} 
This proves \eqref{equchaexpshomin}. \finproof \\
 
We now introduce the
functions 
$\Delta_{F}$ and
$\Gamma_{F}:[a,b]\times F^{-1}(\{\alpha\})\to [0,\infty)$ 
defined by
\begin{align}
&\label{equdefdel}
    \Delta_{F}(u,\theq)\eqd   es(u)-es(q),\\
&\label{equdefgammf}
    \Gamma_{F}(\theu,\theq)\eqd  \frac{\Delta_{F}(\theu,\theq)}{(\theu-\theq)^{2}}\textbf{1}_{(0,\infty)}(|\theu-\theq|)+\frac{\den (\theq)}{2(1-\alpha)}\textbf{1}_{\{0\}}(\theu-\theq),
\end{align}
where  $\den (\theq)$ is the density of $Y$ at $q$ (assumed to exist).
\bl \label{}
Under Assumption \ref{assconfarvar}, if $F$ is absolutely continuous in $(a,b)$ with associated density denoted by $\den$, then  
\begin{align}
\label{equfactdelgam}
&\Delta_{F}(\theu,\theq)= (\theu-\theq)^{2}\Gamma_{{F}}(\theu,\theq)\sp (\theu,\theq)\in[a,b]\times F^{-1}(\{\alpha\})\\
\label{equprogam}
&\Gamma_{{F}}^{-1}(\{0\})\subseteq   F^{-1}(\{\alpha\})\times F^{-1}(\{\alpha\})=\Delta_{F}^{-1}(\{0\})\\
\label{equboudiflos}
 &  \inf_{t\in (0,1)}\den (t\theu+(1-t)\theq)\leq   2(1- \alpha)\Gamma_{F}(\theu,\theq)\leq\sup_{t\in (0,1)}\den (t\theu+(1-t)\theq).
\end{align}

\el
\begin{proof}
    \eqref{equfactdelgam} is immediate from the definition of $\Gamma_{F}$. It is also clear from \eqref{equchavarmin} that $\Delta_{F}$ as in \eqref{equdefdel} is nonnegative, hence so is in turn \eqref{equdefgammf}.
\eqref{equchavarmin} implies at once the equality of sets at the right in \eqref{equprogam}, while the inclusion of sets at the left in \eqref{equprogam} is an immediate consequence of the definition of $\Gamma_{F}$ and of the first equality in \eqref{equchavarmin}.

If $F$ is differentiable in $[a,b]$, 
then \eqref{equchavarmin} implies that for every $(\theu,\theq)\in [a,b]\times F^{-1}(\{\alpha\})$,
\bel
 &   \Delta_{F}(\theq,\theq)=0= \partial_1 \Delta_{F} (\theq,\theq)\sp  \partial^{2}_{ 1,1 } \Delta_{F}  (tu+(1-t)q,\theq)= (1-\alpha)^{-1} \den (t\theu +(1-t)\theq),
\eel
where $\partial_1 \Delta_{F}$ and
$\partial^{2}_{ 1,1 }\Delta_{F}$ refer to the first and second derivatives of $\Delta_F$ with respect to its first argument.
Hence, by an application of Taylor's theorem with the Lagrange form of the remainder, given $(\theu,\theq)\in[a,b]\times F^{-1}(\{\alpha\})$ there exists $t_{\theu,\theq}\in (0,1)$ for which
\begin{align}
\label{equequlamtay}
    \Delta_{F}(\theu,\theq)=\frac{(\theu-\theq)^{2}}{2(1-\alpha)}\den (t_{\theu,\theq}\theq+(1-t_{\theu,\theq}u)).
\end{align}
\eqref{equboudiflos} is then clear from \eqref{equequlamtay}. \finproof\\
\end{proof}

\noindent
In view of \eqref{equdefdel}-\eqref{equdefgammf},  
\eqref{equboudiflos} reflects the connection between the value of $\den$ and the accuracy of $\theu$ as an approximation of $\theq$ as measured by $\Delta_{F}$.

\section{Proofs of Main Results}
\label{sec:tecpro}
 
 \label{sec:prosec:prosetupconvares}

 \subsection{Proof of Lemma \ref{corexifunvarexpsho}}
 \label{secpromeavar}

The functions $\omega\mapsto\VaRa (Y|X(\omega))$ and $\omega\mapsto \ESa (Y|X(\omega))$ are $\sigma(X)$  measurable.  
In fact, given $t\in \R$, we have
\begin{align}
\{\VaRa (Y|X)<t\}=\cup_{n\in \N}\{F_{Y|X}(t-1/n)\geq \alpha\},
\end{align}
which is a countable union of $\sigma(X)$  measurable sets (as $F_{Y|X}(y)$ is $\sigma(X)$  measurable for every fixed $y$). This shows the claim for $\VaRa (Y|X)$.

As for the $\sigma(X)$  measurability of 
$\ESa (Y|X)$, notice that the function $es:\cX \times \R\to \R$ defined by 
\begin{align}
 es(x,\theu)= \frac{1}{1-\mu(x,(-\infty,\theu))}\int y\textbf{1}_{[\theu,\infty)}(y) \,\mu(x,dy)
\end{align}
is Borel measurable (on the set where $\mu(x,(-\infty,\theu))<1$) and that
\begin{align}
\ESa (Y|X)=es(X,\VaRa (Y|X)).
\end{align}
The result then follows by an application of the Doob-Dynkin lemma. \finproof

\subsection{Proof of  Lemma \ref{lemma:bounds:q}}
\label{secprolembouq}

Combining  $\PPro{Y < q(X)|X}\leq {\alpha\leq \PPro{Y \le q(X)|X}}$ with easy inequalities, we get
\begin{align*}
q(X)\geq 0
&\Longrightarrow
(1-\alpha) \, q(X)
\leq q(X)\,\PPro{Y\ge q(X)|X}
\leq \Esp{Y\ind_{Y\ge q(X)}|X}
\leq 
\Esp{Y^+|X},
\\[2mm]
q(X)\leq 0&\Longrightarrow
\alpha\, q(X)\geq 
q(X)\, \PPro{Y{\le }q(X)|X}
\geq 
\Esp{Y\ind_{Y{\le } q(X)}|X}
\geq 
-\Esp{Y^-|X},
\end{align*}
which implies \eqref{bounds:on:VaR}.

The inequality \eqref{bounds:on:ES} stems from the bound $|\ESa (Y|X)|\leq (1-\alpha)^{-1}\Esp{|Y|\,|\,X}$ that follows from Definition \ref{defqualevalp}. These inequalities (resp.\ and the conditional Jensen inequality) show that the (resp.\  square) integrability of $\theq(X)$ and $\thes(X)$ follows from that of $Y$. \finproof

\emmi{add a similar bound for $s(Y)$}

\subsection{Proof of Lemma \ref{theapprisexiden}}
\label{secprolemuppboubia}

For any $f\in\cLab$,  \eqref{equirrwei}  is a consequence of the definition of $\gamma_{f}$ in \eqref{equdefweifuncas} and \eqref{equprogam}.
Now notice that, according to \eqref{equchacondisfunexp}, the definition \eqref{equdefweifuncas} of $\gamma_{f}$, \eqref{equdefdel} and \eqref{equdefinslosvar}, 
 \begin{align}
   \lmeavar(f)-\lmeavar(\theq)=\Esp{\Esp{\lpvar(Y,f(X))-\lpvar(Y,\theq(X))|X}}=\Esp{\Delta_{F_{Y|X}}(f(X),\theq(X))} 
 \end{align}
 for $f\in \cFab$, which leads via \eqref{equfactdelgam} to the equality
 \begin{align}\label{e:leq}
     \lmeavar(f)-\lmeavar(\theq)=\Esp{\gamma_{{f}}(X)(f(X)-\theq(X))^{2}}.
 \end{align}
 Together with 
 \begin{align}
\label{equprodeflemavar}
 \lmeavar(\appvar)-\lmeavar(\theq)\leq \lmeavar(f)-\lmeavar(\theq)   
 \end{align}
   valid for every $f\in \cFab$, this gives rise to 
  \begin{align}
\|\gamma_{{\appvar}}(\appvar-\objfunvar)\|_{\PP_{X},2}^{2}={\lmeavar(\appvar)-\lmeavar(\theq)=  \inf_{f\in \cFab}\|\gamma_{{f}}(f-\objfunvar)\|_{\PP_{X},2}^{2}},
\label{equappvartruvarinel2}
\end{align}
 which implies the equalities in \eqref{equineapperrvar}. In addition, the upper bound
\beql{equappvartruvarinel1}
\lmeavar(\appvar)-\lmeavar(\theq) &= \frac{1}{1-\alpha}\Esp{(Y- \appvar(X))^+ - (Y- \theq(X))^+}+ \Esp{\appvar(X)- \theq(X)}\\
 &\leq \left(\frac{2-\alpha}{1-\alpha}\right)\inf_{f\in \cFab}\|f-\objfunvar\|_{\PP_{X},1}
\eeql
follows via an elementary estimation using 
\begin{align}
 \label{equdifpospardis}
 |w^{+}-v^{+}|\leq |w-v|
 \end{align}
 and the triangle inequality, together with a new application of \eqref{equprodeflemavar}. This shows the inequality at the end in \eqref{equineapperrvar}. 
\finproof 

\subsection{Proof of Theorem \ref{theeststaerrvar}}
\label{secprolemstaerrvar}

{\def \th{\tilde{h}}
\def \hh{\hat{h}_n}\def \hh{\hat{h}}
\def \tR{\tilde{R}}\def \tR{\phi}
\def \hR{\hat{R}_n}\def \hR{\hat{R}}\def \hR{\hat{\phi}}

\def \th{\tilde{p}}
\def \hh{\hat{h}_n}\def \hh{\hat{p}}
\def \tR{\tilde{R}}\def \tR{\Theta}
\def \hR{\hat{R}_n}\def \hR{\hat{R}}\def \hR{\widehat{\Theta}}
Before proving the theorem  we introduce the following Rademacher bound.
\bl\label{lem:rade} Let $Z_{\nn}  $ be an i.i.d.\ sample of a random element $Z$ of $\theP$, with $Z$ independent of $Z_{\nn}  $,  and let $\cH$ be a family of functions $\theP\to \R$. Define, for $z\in \theP$,
$
A_{\cH}(z)\eqd  \inf_{h\in \cH} h(z),  B_{\cH}(z)\eqd  \sup_{h\in \cH} h(z), 
$
and assume that 
$
0<\|B_{\cH}- A_{\cH}\|_{\PP_{Z}, \infty} <\infty.
$
For a possibly  data dependent $h\in \cH$, let
$
\tR(h) = \Esp{h(Z ) | Z_{\nn}},$ $\hR(h) = \frac{1}{n}\sum_{i =1}^n h(Z_k),
$
$\th\in \argmin  _{h\in \cH}\tR(h)$, $\hh\in \argmin  _{h\in \cH}\hR(h)\label{e:minh}.
$
Then for any $\delta\in (0,1)$ 
\beql{e:lemra}
\tR(\hh) - \tR(\th) \leq 2\Radcave(\cH, Z_{\nn}  ) + \|B_{\cH}- A_{\cH}\|_{\PP_Z, \infty} \sqrt{\frac{2\log\frac{2}{\delta}}{n}}
\eeql
holds with probability at least $1-\delta$.
\el
\proof First,
\beql{e:ra}
\tR(\hh) - \tR(\th) &= \Big(\tR(\hh) - \hR(\hh)\Big) + \Big(\hR(\hh) - \hR(\th)\Big) + \Big(\hR(\th) - \tR(\th)\Big)\\
&\leq \Big(\tR(\hh) - \hR(\hh)\Big) + \Big(\hR(\th) - \tR(\th)\Big).
\eeql
To bound from above the first difference at the second line in \eqref{e:ra}, we define the family $\cH'$ by
\begin{align}
    \cH' \eqd \frac{\cH -A_{\cH}}{\|B_{\cH}- A_{\cH}\|_{\PP_Z, \infty}}\eqd  \left\{\theP \ni z \mapsto\frac{h(z)-A_{\cH}(z)}{ \|B_{\cH}- A_{\cH}\|_{\PP_Z, \infty} };\; h\in \cH  \right\}.
\end{align}
It then follows from  Definition \ref{defradcom} that
\beql{e:ra1}
 &\Radcemp(\cH', Z_{\nn}  ) = \Radcemp\left(\frac{\cH -A_{\cH}}{\|B_{\cH}- A_{\cH}\|_{\PP_Z, \infty}}, Z_{\nn}  \right)\\
  &\qqq=\frac{\Radcemp(\cH, Z_{\nn}  ) + \Radcemp({A_\cH}, Z_{\nn}  )}{\|B_\cH - A_\cH\|_{\PP_Z, \infty}}  = \frac{\Radcemp(\cH, Z_{\nn}  )}{\|B_\cH - A_\cH\|_{\PP_Z, \infty}},
\eeql
where the last equality follows from the fact that the Rademacher complexity of a single function set is null. By \citet*[Eqn.\ (3.7) and (3.13) p.31--32]{mohri2018foundations:2nd:edition}, 
\beql{e:ra2}
\sup_{h\in \cH'} \left(\tR(h)- \hR(h)\right)\leq 2\Radcave(\cH', Z_{\nn}  ) + \sqrt{\frac{\log\frac{2}{\delta}}{2n}}
\eeql
holds with probability at least $1-\delta/2$.
Multiplying both sides of the above inequality by $\|B_\cH - A_\cH\|_{\PP_Z, \infty}$, changing the set of the sup term in the left-hand side to $\cH'$ and using \eqref{e:ra1} yields 
\beql{e:ra3}
\tR(\hh) - \hR(\hh) \leq \sup_{h\in \cH} \left(\tR(h)- \hR(h)\right) & \leq 2\Radcave(\cH, Z_{\nn}  ) + \|B_\cH - A_\cH\|_{\PP_Z, \infty}\sqrt{\frac{\log\frac{2}{\delta}}{2n}}  
\eeql
with probability at least $1-\delta/2$. 
Since $\th$ does not depend on $Z_{\nn}  $, the
difference $\hR(\th) - \tR(\th)$ in \eqref{e:ra} can be bounded using the same argument with $\cH=\{\th\}$ to obtain
\beql{e:ra4}
\hR(\th) - \tR(\th)\leq  \|B_\cH - A_\cH\|_{\PP_Z, \infty}\sqrt{\frac{\log\frac{2}{\delta}}{2n}}
\eeql
with probability at least $1-\delta/2$.
Combining 
\eqref{e:ra},
\eqref{e:ra3} and \eqref{e:ra4} yields \eqref{e:lemra}.\ \finproof \\
 
Back to the proof of the theorem, applying \eqref{e:leq} (valid for every $f\in \cFab$) to $f=\empappvar\in \cFab$ yields
\begin{align}
\|(\empappvar-\theq)\gqh\|_{\PP_{X},2}^{2} = \lmeavar(\empappvar) - \lmeavar(\theq) =  \left(\lmeavar(\empappvar)-\lmeavar(\appvar)\right)+\big(\lmeavar(\appvar)-\lmeavar(\theq)\big).\label{equradestconintapp:0}
\end{align}
The term $(\lmeavar(\appvar)-\lmeavar(\theq))$ is handled by \eqref{equineapperrvar}.  To upper bound $\lmeavar(\empappvar)-\lmeavar(\appvar)$ in probability, notice that
\begin{align}
\label{equrewlpvar}
    (1-\alpha)\lpvar(y,u)=&(y-\alpha u)\lor ((1-\alpha)u).
\end{align}
If both $y$ and $u$ are in the range $( v,w)$ for $  \infty< v\leq w <\infty$, then a little algebra shows,  via \eqref{equrewlpvar}, that
\beql{e:brho0}
    (1-\alpha)v \leq(1-\alpha)\lpvar(y,u)\leq w-\alpha v,
\eeql
which implies that   
\beql{e:brho}
    a(X) \leq\lpvar(\Yab,f(X))\leq (1-\alpha)^{-1}(b(X)-\alpha a(X))\sp \mbox{$\PP$  a.s.}.
\eeql
Hence, an application of Lemma \ref{lem:rade}  with $Z_{\nn}   = (X,\Yab)_{\nn}   $, $\cH =\lpvar({\cFab}) \eqd  \{\cX\times \R \ni(x,y)\mapsto\lpvar(y,f(x)); \; f\in \cFab\}$, $A_\cH = a$ and $B_\cH =(1-\alpha)^{-1} (b-\alpha a)$ results in the inequality
\beql{equradestconintapp}
&\lmeavar(\empappvar)-\lmeavar(\appvar)\leq 2\Radcave(\lpvar({\cFab}),{(X,\Yab)_{\nn}  }) 
 + \frac{\|b- a\|_{\PP_X,\infty}\sqrt{2\log({2}/{\delta})}}{(1-\alpha)\sqrt{n}}
\eeql
with probability at least $1-\delta$. 

Note  now that, owing to Talagrand contraction lemma (\citet[Lemma 5.7 p.93]{mohri2018foundations:2nd:edition}), since 
$u\mapsto (1-\alpha)^{-1} u^{+}$
is $(1-\alpha)^{-1}$  Lipschitz, we have for any $(x,y)_{\nn}  \subset({\cX}\times\R)^{n}$:
\beql{e:rade_lips}
\Radcemp(\lpvar(\cFab),(x,y)_{\nn}  )\leq& \Radcemp(\{(1-\alpha)^{-1}(y-f)^{+}:f\in \cFab\},(x,y)_{\nn}  )+\Radcemp(\cFab,x_{\nn}  )\\
\leq & (1-\alpha)^{-1}\Radcemp(\{y-f:f\in \cFab\},(x,y)_{\nn}  )+\Radcemp(\cFab,x_{\nn}  )\\
\leq& (1-\alpha)^{-1}\Radcemp(\{y\},y_{\nn}  )+\left(\frac{2-\alpha}{1-\alpha}\right)\Radcemp(\cFab,x_{\nn}  )\\
=&\left(\frac{2-\alpha}{1-\alpha}\right)\Radcemp(\cFab,x_{\nn}  ),
\eeql
where the last equality follows from Definition \ref{defradcom}. The corresponding inequality for average complexities,
\begin{align}
\label{equineradcomlosvar}\Radcave(\lpvar(\cFab), (X,\Yab)_{\nn}  )\leq& \left(\frac{2-\alpha}{1-\alpha}\right)\Radcave(\cFab,X_{\nn}  ),
\end{align}
follows by integration with respect to the law of $(X,Y)_{\nn}  $. A combination of \eqref{equradestconintapp:0} with \eqref{equradestconintapp},    \eqref{equineradcomlosvar}, \eqref{equineapperrvar} and  the inequality 
$
\sqrt{w^2+v^2}\leq |w|+|v|
$ yields
 \eqref{equesterrvardel}. 
 \finproof
}

\subsection{Massart's Lemma}
\bl
\label{lem:mas} Let $Z_{\nn}  $ be a sequence of random variables, and consider  a family of functions $\cH$ mapping $\theP$ to $\R$
, then 
\begin{align}
\label{e:mass}
n\Radcave(\cH, Z_{\nn}  ) \leq& \varepsilon +   \Esp{\sup_{h\in \cH}\sqrt{2\log (\cN_{1}(\cH,Z_{\nn}  ,\varepsilon/{n}) )\sum_{i =1}^n h^2(Z_i)  }}\\
\leq& \varepsilon + \sqrt{2n}\,
||\sup_{h\in \cH}|h(Z_1)|\,||_{\PP,\infty}\Esp{\sqrt{\log (\cN_{1}(\cH,Z_{\nn}  ,\varepsilon/{n}) )}}
.
\end{align}
holds for any $ \varepsilon>0$.
\el
\proof Integrating the estimate in \citet*[Proposition 5.2]{rebe2021} yields the first inequality, from which the second one is deduced by upper bounding the Euclidean norm of $(h(Z_1),\dots ,h(Z_n))$ with $\sqrt{n}||\sup_{h\in \cH}|h(Z_1)|\,||_{\PP,\infty}$. \finproof

\subsection{Proof of Theorem \ref{thebouerrrrad}} \label{secprolemristofun}

We  consider the family of functions $\lpesB _{f}(\cGB)$ defined (for fixed  $f$) by
\begin{align}
\lpesB _{f}(\cGB)\eqd \{\cX\times\R\ni(x,y)\mapsto \lpesB (y,f(x),g(x))\in\R
;\, g\in \cGB\}.
\end{align} 
We also denote by $ \ther_{f}$ any function
$$
    \ther_{f} \in \argmin  _{g\in\cLp \nnx  } 
    \Esp{
(\lpvar_{\hun}(Y,f(X)  )-f(X)-g(X) )^{2}
}  
,$$
i.e., as $(Y-f(X))^{+}$ is square integrable, any function $\ther_{f} \in\cLp \nnx  $ such that   
\begin{align}
\ther_{f}(X)=\Esp{(1-\alpha)^{-1}(Y-f(X))^{+}|X},\qq \PP\mbox{ a.s.},
\end{align}
and we let $\therB_{f}: \cX \to [0,B] $ be one of its truncated companions, in the sense that
\begin{align}
\therB_{f}(X)=\Esp{\Tb{(1-\alpha)^{-1}(Y-f(X))^{+}}|X},\qq \PP\mbox{ a.s.}.
\end{align}

Let now $\|\cdot\|$ be a shorthand for the $L^{2}_{\PP_{X}}$ seminorm on the space of square integrable measurable functions on $\cX$.

\bl
\label{lemdifloseslosvar}
For every  square integrable functions $f,f'$, we have
\begin{align}
&\|\ther_{f}-\therB_{f}\|\leq \|((1-\alpha)^{-1}(Y-f(X))^{+}-B)^{+}\|_{\PP,2},\\
&\|\therB_{f}-\therB_{f'}\| \leq (1-\alpha)^{-2}\|f-f'\|.
\label{inerfminrvar}
\end{align} 
\el

\begin{proof}
The first inequality is a direct consequence of Jensen's inequality 
\begin{align}
\Esp{|\Esp{(W-\Tbb W)|X}|^{2}}\leq 
\Esp{|W-\Tbb W|^{2}}=
\Esp{((W-B)^{+})^{2}},
\end{align}
valid for any square integrable {positive} random variable $W$. As for the second one, notice that $\Tbb\cdot $ is 1-Lipschitz.
 Combining this property with \eqref{equdifpospardis} and with Jensen's inequality we obtain:
\begin{align*}
\|\therB_{f}-\therB_{f'}\|^{2}
&\leq \Esp{\Esp{\left|\Tb{\frac{(Y-f(X))^{+}}{(1-\alpha)}}-\Tb{\frac{(Y-f'(X))^{+}}{(1-\alpha)}}\right|^{2}\Bigg|X}}\\
 & \leq(1-\alpha)^{-2}\|f-f'\|^{2}.~\finproof\fe
\end{align*}
\end{proof}

Back to the proof of Theorem \ref{thebouerrrrad},  for $f\in \cLabt$, the triangle inequality 
gives
\begin{align}
\|\empappexpshoB{f}-\ther\|\leq& \|\empappexpsho{f}-\therB_{f}\|+\|\therB_{f}-\therB_{\theq}\|+\|\therB_{q}-\ther\|\\
\leq &\|\empappexpsho{f}-\therB_{f}\|
+(1-\alpha)^{-1}\|f-\theq\|+\|((1-\alpha)^{-1}(Y -\theq(X))^{+}-B)^{+}\|_{\PP,2},\qquad
\label{equprochaava}
\end{align}
by Lemma \ref{lemdifloseslosvar}.

For a fixed $f$, applying Lemma \ref{lem:rade} to $Z_{\nn}   = (X,Y)_{\nn}   $, $\cH = \lpesB _{f}(\cGB)$, and to the constant functions $A_\cH=0$ and $ B_\cH = \sup_{g\in \cGB}\|
\lpesB (Y,f(X),g(X))\|_{\PP,\infty}$  yields
\beql{equradestconintfores}
&\|\empappexpshoB{f}-\therB_{f}\|%
^{2}-\inf_{g\in \cGB}\|g-\therB_{f}\|%
^{2}\\ 
\leq&2\Radcave(\lpesB _{f}(\cGB),(X,Y)_{\nn})  + \frac{\sup_{g\in \cGB}\|\lpesB (Y,f(X),g(X))\|_{\PP,\infty}\sqrt{2\log({2}/{\delta})}}{\sqrt{n}}  
\eeql
with probability at least $1-\delta$, for any given $\delta\in (0,1)$. 
In addition, Lemma \ref{lemdifloseslosvar} together with  the triangle inequality
implies that 
\beql{egb}
\inf_{g\in \cGB}\|g-\therB_{f}\| \leq \inf_{g\in \cGB}\|g-\ther\| +(1-\alpha)^{-1}\|f-\theq\|+\|((1-\alpha)^{-1}(Y -\theq(X))^{+}-B)^{+}\|_{\PP,2}.
 \eeql
In virtue of the triangle inequality and the inequality $\sqrt{w
^{2}+v^{2}}\leq |w|+|v|$, replacing \eqref{egb} in \eqref{equradestconintfores} and combining the resulting inequality with \eqref{equprochaava} implies 
\beql{equradestconintforesproalm}
&\|\empappexpshoB{f}-\ther\|
\leq  \inf_{g\in \cGB}\|g-\ther\|
\\
&\quad+ 2 \Big((1-\alpha)^{-1}\|f-\theq\|
+\|((1-\alpha)^{-1}(Y -\theq(X))^{+}-B)^{+}\|_{\PP,2}\Big)
\\
&\quad+\left(2\Radcave(\lpesB _{f}(\cGB),(X,Y)_{\nn}  + \frac{\sup_{g\in \cGB}\|\lpesB (Y,f(X),g(X))\|_{\PP,\infty}\sqrt{2\log({2}/{\delta})}}{\sqrt{n}}\right)^{1/2}.
\eeql
Let us now upper bound the Rademacher complexity in the above equation. 
Since the square function on $[\eg{-B},B]$ 
has Lipschitz constant equal to $2B$, Talagrand's contraction lemma
 gives
\begin{align}
\Radcave(\lpesB _{f}(\cGB),(X,Y)_{\nn}  )
\leq 2B \Radcave(\cGB,X_{\nn}  ).
\label{equrearad}
\end{align}
Replacing \eqref{equrearad} and the bound
$\sup_{g\in \cGB}\|\lpesB _{f}(g)(X,Y)\|_{\PP,\infty}\leq  B^{2}$ in \eqref{equradestconintforesproalm} implies \eqref{equesterrvardelentestes}. \finproof

 
\subsection{Proof of Proposition \ref{p:twin}}\label{ss:prooftwin}
We have
\[
\|\PPro{Y> \thisq (X)|X}-1+\alpha\|^2_{\PP ,2} =  \Esp{\PPro{Y> \thisq (X)|X}^2}+(1-\alpha)^2-2(1-\alpha) \PPro{Y> \thisq (X)}  ,
\]
where
\[  \PPro{Y> \thisq (X)|X}^2
= \PPro{Y^{(1)}\wedge Y^{(2)}> \thisq (X)|X}.\]
Thus \eqref{eq:twinsim_var_err} follows.
For the ES, we have $\mathrm{E}[\lpvar (Y, q (X))|X]= s(X)$ (see \eqref{equdefinslosvar} and \eqref{equexpshoconexp}), hence
\begin{align*}&\|\this(X)-s(X) \|^2_{\PP ,2} = 
\|\Esp{Z|X} \|^2_{\PP ,2},\end{align*}
where $Z\eqd \this(X)-\lpvar (Y, q (X))$ satisfies by the conditional Jensen inequality:
\beql{e:jens}&\|\Esp{Z|X} \|^2_{\PP ,2}=\Esp{(\Esp{ Z|X} )^2} \le \Esp{\Esp{ Z^2|X}} = \Esp{Z^2}=
\|\this(X)-\lpvar(Y,q(X))\|^2_{\PP ,2} .
\eeql
An application of the triangular inequality yields
\begin{align*}&\|\this(X)-\lpvar(Y,q(X))\|_{\PP ,2} \leq%
 \|\this(X)-\lpvar(Y,\thisq (X))\|_{\PP ,2}+\|\lpvar(Y,q(X))-\lpvar(Y,\thisq (X))\|_{\PP ,2}.\end{align*}

By $\frac{2-\alpha}{1-\alpha}$  Lipschitz regularity of $\lpvar$ with respect to its second argument,
\begin{align}&\|\lpvar(Y,q(X))-\lpvar(Y,\thisq (X))\|_{\PP ,2}
\leq%
\frac{2-\alpha}{1-\alpha}\|q(X)- \thisq (X)\|_{\PP ,2}.\label{e:lip}\end{align}
Given our assumption that
 $\inf_{y\in (a(X),b(X))} \dot{F}_{Y|X }(y) \geq c$  holds $\PP$ a.s.,
 it follows that, $\PP$ a.s., $$\sup_{t\in \big(\alpha\wedge( 1-\PP [Y > \thisq (X)|X])),\alpha\vee (1-\PP [Y > \thisq (X)|X])\big) }  
 \dot{\wideparen{F_{Y|X}^{-1}}} (t)
 \leq \sup_{t\in \big(F^{-1}(a(X)), F^{-1}(b(X))\big)} \dot{\wideparen{F_{Y|X}^{-1}}} (t)
\leq \frac{1}{c} .$$
By writing $q(X) = F_{Y|X}^{-1}( \alpha)$ and $f(X) = F_{Y|X}^{-1}(1-\PP [Y > \thisq (X)|X])$, applying the mean value theorem, we obtain
\bel
| q(X)- \thisq (X)|=|\alpha-1+\PP [Y > \thisq (X)|X]|\dot{\wideparen{F_{Y|X}^{-1}}} (t) 
\eel
for some $t\in \big(\alpha\wedge( 1-\PP [Y > \thisq (X)|X])),\alpha\vee (1-\PP [Y > \thisq (X)|X])\big)$, $\PP$ a.s.. Hence 
\bel
\|q(X)- \thisq (X)\|_{\PP ,2} \leq \frac{1}{c}\|1-\alpha - \PP [Y > \thisq (X)|X]\|_{\PP ,2}.
\eel
Replacing the above bound to \eqref{e:lip} yields
\eqref{eq:es_err_decompo}. 
By conditional independence and tower law, we get \eqref{eq:twinsim_es_err_proxy}. \finproof

\subsection{Proof of Lemma \ref{l:radenn}\label{s:bnn}}
 For any $x \in \cX$, $\tx = \begin{bmatrix} x \\1 \end{bmatrix}$,  and $f \in \cNN(d, 1,B_{1:(l+1)},l,m,\sigma)$, an application of \eqref{mulprofro}
 yields
 \beql{e:cau}
 |f(x)| = | W_{l+1}\sigma(W_{l}\sigma(\dots\sigma(W_1 \tx))| \leq  | W_{l+1}|_2 \, |\sigma(W_{l}\sigma(\dots\sigma(W_1 \tx))|_2.
 \eeql
 Since $\sigma$ is 1-Lipschitz and positive homogeneous, $
 |\sigma(V)|_2 \leq  |V|_2 
$ holds for any vector $V$.
 Successive applications of this and of \eqref{mulprofro} imply \eqref{e:bound_nn}.

Given \eqref{e:bound_nn},
integrating both sides of the inequality in \citet[Theorem 1]{golowich18a} w.r.t. the law of $X_{\nn}  $ yields
\eqref{e:rade_nn_relu}. \finproof

\subsection{Proof of Theorem \ref{t:cvrateNNq} }
\label{s:proof_nnvar}
By Lemma \ref{l:radenn}, $\cFab = \cNN(d,1,B_{1:(l+1)},l,m,\sigma)$ is uniformly bounded and $ \cFab \subseteq \cLabt$. 
Similar computations as in Section \ref{s:bnn} show that for $f\in\cFab$,  $x, x' \in \cX$ and $\delta>0$, if $|x  -x'|_2 <\delta$, then
    \beql{}
    |f(x ) - f(x')|_2\leq \prod_{k=1}^{l+1}B_k \delta\sp  f\in \cFab,
    \eeql
    which implies the equicontinuity of $\cFab$. By the Ascoli-Arzel\`a theorem that then applies under the assumption {\em (i)} (see e.g.\ \citet[Theorem 4.25 p.111]{brezis2011functional}), 
    $\cFab$ is compact in the space of continuous functions on $\cX$. Now let $\{f_j\}$ be a sequence  in $\cFab$ that converges pointwise to $f\in \cFab$. Then the same calculations  as for \eqref{e:brho} yield, for any $j$,
    \beql{}
    |\lpvar(Y^{a,b},f_j(X))| \leq a(X) \wedge (1-\alpha)^{-1}\Big(b(X)-\alpha a(X)\Big),
    \eeql
    which is integrable by assumption \textit{(i)}. Hence, by dominated convergence, we have 
    \beql{}
    \lim_{j\to \infty} \lmeavar(f_j) = \lmeavar(f)\sp \mbox{and\;} \lim_{j\to \infty} \ave{n}\lpvar_{\hun}\left(\Yab_i,f_j(X_{i}) \right) = \ave{n}\lpvar_{\hun}\left(\Yab_i,f(X_{i}) \right),
    \eeql
    which implies the continuity of the risk functions. This and the compactness of $\cFab$ results in the existence of $\appvar$ and $\empappvar$ by an application of the Weierstrass extreme value theorem.
    
    By  combining \eqref{equesterrvardel2} with the assumption {\em (ii)} and \eqref{e:rade_nn_relu}, we obtain \eqref{equesterrvardelentestneunet}. \finproof

    \subsection{Proof of Theorem \ref{t:cvesnn}} \label{s:proof_esnn}
By Lemma \ref{l:radenn},  $\cGB$ is uniformly bounded. Analogous reasoning used to prove Theorem \ref{t:cvrateNNq} leads to the compactness of $\cGB$ in the space of continuous functions on $\cX$. Let $g_j$ functions  in $\cGB$  converge pointwise to $g\in \cGB$.  For any fixed $f\in \cL\nnx $, by continuity of $z\mapsto \left(\Tb{(1-\alpha)^{-1}(y- f(x)  )^{+}} - z\right)^2$, we then have that $$
 \lim_{j\to \infty}\ave{n}{ \lpesB _{\hde}(Y_{i},f(X_{i}),g_j(X_{i}))}  =  
\ave{n}{ \lpesB _{\hde}(Y_{i},f(X_{i}),g(X_{i}))},
$$
which proves the continuity of the risk  
 $g\mapsto\lempestrunc_f (g )$. The existence of $\empappexpsho{f%
}$ follows by an application of the Weierstrass extreme value theorem. \finproof

\subsection{Proof of the Toy Model ES Formula (\ref{e:ESt})\label{e:sk}}
By \citet[Eqn.\ (5c)]{khokhlov2016conditional}, one has
\begin{align} 
    y\den _\nu(y, \mu,\sigma )  = \partial_y \left(-\frac{\nu\sigma^2+ (y-\mu)^2}{\nu -1}\den _\nu(y, \mu,\sigma ) +\mu \Upsilon_\nu(y,\mu,\sigma )\right). 
\end{align}
When $y$ tends to $\infty$, the cdf $\Upsilon_\nu(y,\mu,\sigma )$ tends to 1 and, for $\nu >1$, $\frac{\nu\sigma^2+ (y-\mu)^2}{\nu -1}\den _\nu(y, \mu,\sigma ) $ tends to 0, because the pdf converges to 0 faster than the polynomial of order 2 goes to $\infty$. Hence
\begin{align} \label{eq:tt1}
     \int_{y_0}^\infty y \den _\nu(y, \mu,\sigma ) dy = \mu(1- \Upsilon_\nu(y_0,\mu,\sigma ) ) + \frac{\nu\sigma^2+ (y_0-\mu)^2}{\nu -1}\den _\nu(y_0, \mu,\sigma ). 
\end{align}
For conciseness, in the following calculus, we use $y_0 = {\Upsilon}^{-1}_{Y|X} (\alpha) $ and functions $P,Q,S$ without the input.  By \eqref{eq:t_pdf} and \eqref{eq:tt1},
\begin{align*}
    &\ES(Y|X)= \frac{1}{1-\alpha}\int_{y_0}^\infty y  {\den }_{Y|X}(y) d y \\
    &= \frac{1}{(1-\alpha)}\int_{y_0}^\infty \frac{y}{2} \left[ \den _\nu(y, P-Q, S) +  \den _\nu(y, P+Q, S) \right] d y \\
    &= \frac{1}{2(1-\alpha)} \bigg[(P-Q)(1-\Upsilon_\nu(y_0, P-Q, S))+  \frac{\nu S^2+ (y_0-P+Q)^2}{\nu -1}\den _\nu(y_0, P-Q, S )  \\
    &\qq+ (P+Q)(1-\Upsilon_\nu(y_0, P+Q, S))+  \frac{\nu S^2+ (y_0-P-Q)^2}{\nu -1}\den _\nu(y_0, P+Q, S )\bigg]\\
    &=
    \frac{1}{2(1-\alpha)} \bigg[2P - 2P{\Upsilon}_{Y|X}(y_0)+ Q\times(\Upsilon_\nu(y_0, P-Q, S)-\Upsilon_\nu(y_0, P+Q, S))+ \\
    &\qq2\frac{\nu S^2+ (y_0-P)^2+ Q^2}{\nu -1}{\den }_{Y|X}(y_0 ) +2 \frac{(y_0-P)Q }{\nu -1}(\den _\nu(y_0, P-Q, S )- \den _\nu(y_0, P+Q, S )) \bigg],
\end{align*}
by \eqref{eq:t_cdf} and \eqref{eq:t_pdf} again. Reminding that $y_0 = {\Upsilon}^{-1}_{Y|X} (\alpha) $, we obtain 
\eqref{e:ESt}. \finproof

\bibliographystyle{chicago}
\bibliography{BIB}

\end{document}